\documentclass[format=acmsmall, review=false]{acmart}
\usepackage{acm-ec-21}
\usepackage{booktabs} 
\usepackage[ruled]{algorithm2e} 

\SetAlFnt{\small}
\SetAlCapFnt{\small}
\SetAlCapNameFnt{\small}
\SetAlCapHSkip{0pt}
\IncMargin{-\parindent}

\usepackage{amsmath} 
\usepackage{amsthm} 
\usepackage{mathrsfs}
\usepackage{bbm}
\usepackage{bookmark}
\usepackage{hyperref}
\usepackage{subcaption}

\newtheorem{theorem}{Theorem}

\newtheorem{proposition}{Prop.}

\setcitestyle{acmnumeric}

\title[(In)Stability for the Blockchain]{(In)Stability for the Blockchain: Deleveraging Spirals and Stablecoin Attacks}

\author{Ariah Klages-Mundt}
\affiliation{%
	\institution{Cornell University, Center for Applied Mathematics}
}
\author{Andreea Minca}
\affiliation{%
	\institution{Cornell University, Operations Research \& Information Engineering}
}

\date{\today}

\begin{document}

\begin{abstract}
\begin{center}
	\today \\
	Initial release: June 2019
\end{center}
\vspace{0.2cm}

We develop a model of stable assets, including non-custodial stablecoins backed by cryptocurrencies. Such stablecoins are popular methods for bootstrapping price stability within public blockchain settings. We derive fundamental results about dynamics and liquidity in stablecoin markets, demonstrate that these markets face deleveraging feedback effects that cause illiquidity during crises and exacerbate collateral drawdown, and characterize stable dynamics of the system under particular conditions. The possibility of such `deleveraging spirals' was first predicted in the initial release of our paper in 2019 and later directly observed during the `Black Thursday' crisis in Dai in 2020. From these insights, we suggest design improvements that aim to improve long-term stability. We also introduce new attacks that exploit arbitrage-like opportunities around stablecoin liquidations. Using our model, we demonstrate that these can be profitable. These attacks may induce volatility in the `stable' asset and cause perverse incentives for miners, posing risks to blockchain consensus. A variant of such attacks also later occurred during Black Thursday, taking the form of mempool manipulation to clear Dai liquidation auctions at near zero prices, costing \$8m.
\end{abstract}

\maketitle

\section{Introduction}\label{sec:introduction}

In 2009, Bitcoin \cite{bitcoin2009} introduced a new notion of decentralized cryptocurrency and trustless transaction processing. This is facilitated by blockchain, which introduced a new way for mistrusting agents to cooperate without trusted third parties. This was followed by Ethereum \cite{ethereum2014}, which introduced generalized scripting functionality, allowing `smart contracts' that execute algorithmically in a verifiable and somewhat trustless manner. Cryptocurrencies promise notions of cryptographic security, privacy, incentive alignment, digital usability, and open accessibility while removing most facets of counterparty risk. However, as these cryptocurrencies are, by their nature, unbacked by governments or physical assets, and the technology is quite new and developing, their prices are subject to wild volatility, which affects their usability.

A stablecoin is a cryptocurrency with an economic structure built on top of blockchain that aims to stabilize the purchasing power of the coin. A true stablecoin, often referred to as the ``Holy Grail of crypto'', would offer the benefits of cryptocurrencies without the unusable volatility and remains elusive. A more tangible goal is to design a stablecoin that maximizes the probability of remaining stable long-term. If one can establish guarantees for the stability of such a stablecoin, this would be a significant step toward forming a robust decentralized financial system and facilitating economic adoption of cryptocurrencies.

\paragraph{Cryptocurrency volatility.}
Cryptocurrencies face difficult technological, usability, and regulatory challenges to be successful long-term. Many cryptocurrency systems develop different approaches to solving these problems. Even assuming the space is long-term successful, there is large uncertainty about the long-term value of individual systems.

The value of these systems depends on network effects: value changes in a nonlinear way as new participants join. In concrete terms, the more people who use the system, the more likely it can be used to fulfill a given real world transaction. The success of a cryptocurrency relies on a mass of agents--e.g., consumers, businesses, and/or financial institutions--adopting the system for economic transactions and value storage. Which systems will achieve this adoption is highly uncertainty, and so current cryptocurrency positions are very speculative bets on new technology. Further, cryptocurrency markets face limited liquidity and market manipulation. In addition, the decentralized control and privacy features of cryptocurrencies can be at odds with desires of governments, which introduces further uncertainty around attempted interventions in the space.

These uncertainties drive price volatility, which feeds back into fundamental usability problems. It makes cryptocurrencies unusable as short-term stores of value and means of payment, which increases the barriers to adoption. Indeed, today we see that most cryptocurrency transactions represent speculative investment as opposed to typical economic activity.

\paragraph{Stablecoins.}
Stablecoins aim to bootstrap price stability into cryptocurrencies as a stop-gap measure for adoption. Current projects take one of two forms:
\begin{itemize}
	\item \textbf{Custodial stablecoins} rely on trusted institutions to hold reserve assets off-chain (e.g., \$1 per coin). This introduces counterparty risk that cryptocurrencies otherwise solve.
	\item \textbf{Non-custodial (or decentralized) stablecoins} create on-chain risk transfer markets via complex systems of algorithmic financial contracts backed by volatile cryptoassets.
\end{itemize}
We focus on non-custodial stablecoins and, more generally, the stable asset and risk transfer markets that they represent. Non-custodial systems are not well understood whereas custodial stablecoins can be interpreted using existing well-developed financial literature. Further, non-custodial stablecoins operate in the public/permissionless blockchain setting, in which any agent can participate. In this setting, malicious agents can participate in stablecoin systems. As we will see, this can introduce new economic attacks.

\subsection{Non-custodial (decentralized) stablecoins}

The non-custodial stablecoins that we consider create systems of contracts on-chain with the following features encoded in the protocol. We refer to these as \textbf{DStablecoins}.
\begin{itemize}
	\item Risk is transferred from stablecoin holders to speculators. Stablecoin holders receive a form of price insurance whereas speculators expect a risky return from a leveraged position.\footnote{`Leverage' means that the speculator holds $>1\times$ their initial assets but faces new liabilities.}
	\item Collateral is held in the form of cryptoassets, which backs the stable and risky positions.
	\item An oracle provides pricing information from off-chain markets.
	\item A dynamic deleveraging process balances positions if collateral value deviates too much.
	\item Agents can change their positions through some pre-defined process.
\end{itemize}
These systems are non-custodial (or decentralized) because the contract execution and collateral are all completely on-chain; thus they potentially inherit all of the benefits of cryptocurrencies, such as minimization of counterparty risk. DStablecoins are variants on contracts for difference, which we describe next. The risk transfer typically works by setting up a tranche structure in which losses (or gains) are borne by the speculators and the stablecoin holder holds an instrument like senior debt.\footnote{Intuitively, these are like collateralized debt obligations (CDOs) with the important addition of dynamic deleveraging according to the rules of the protocol. As we will see, it is critical to understand deleveraging spirals as they affect the senior tranches.} There are also other \emph{non-collateralized} (or \emph{algorithmic}) stablecoins--for a discussion of these, see \cite{blockchain2019}. We don't consider these directly in this paper; however, we discuss in Section~\ref{sec:discussion} how our model can accommodate these systems as well.

\paragraph{Contract for difference.} Two parties enter an overcollateralized contract, in which the speculator pays the buyer the difference (possibly negative) between the current value of a risky asset and its value at contract termination.\footnote{Intuitively, this is similar to a forward contract except that the price is only fixed in fiat terms while payout is in the units of the underlying collateral.} For example, a buyer might enter 1 Ether into the contract and a speculator might enter 1 Ether as collateral. At termination, the contract Ether is used to pay the buyer the original dollar value of the 1 Ether at the time of entry. Any excess goes to the speculator. If the contract approaches undercollateralization (if Ether price plummets), the buyer can trigger early settlement or the speculator can add more collateral.

\paragraph{Variants on contracts for difference.}
DStablecoins differ from basic contracts for difference in that (1) the contracts are multi-period and agents can change their positions over time, (2) the positions are dynamically deleveraged according to the protocol, and (3) settlement times are random and dependent on the protocol and agent decisions. The typical mechanics of these contracts are as follows:
\begin{itemize}
	\item Speculators lock cryptoassets in a smart contract, after which they can create new stablecoins as liabilities against their collateral up to a threshold. These stablecoins are sold to stablecoin holders for additional cryptoassets, thus leveraging their positions.
	
	\item At any time, if the collateralization threshold is surpassed, the system attempts to liquidate the speculator's collateral to repurchase stablecoins/reduce leverage.
	
	\item The stablecoin price target is provided by an oracle. The target is maintained by a dynamic coin supply based on an `arbitrage' idea. Notably, this is not true arbitrage as it is based on assumptions about the future value of the collateral.
	\begin{itemize}
		\item If price is above target, speculators have increased incentive to create new coins and sell them at the `premium price'.
		
		\item If price is below target, speculators have increased incentive to repurchase coins (reducing supply) to decrease leverage `at a discount'.
	\end{itemize}
	
	\item Stablecoins are redeemable for collateral through some process. This can take the form of global settlement, in which stakeholders can vote to liquidate the entire system, or direct redemption for individual coins. Settlement can take 24 hours-1 week.
	
	\item Additionally, the system may be able to sell new ownership/decision-making shares as a last attempt to recapitalize a failing system -- e.g., the role of MKR in \href{https://makerdao.com/dai}{Dai} (see \cite{dai_white2017}).
\end{itemize}

\paragraph{DStablecoin risks.}
DStablecoins face two substantial risks:
\begin{enumerate}
	\item Risk of market collapse,
	\item Oracle/governance manipulation.
\end{enumerate}
Our model in this paper focuses on market collapse risk. We further remark on oracle/governance manipulation in Section~\ref{sec:discussion}.

\paragraph{Existing DStablecoins.}
At the time of initial writing in 2019, major non-custodial stablecoins included Dai, BitShares Market Pegged Assets (like bitUSD), and Steem Dollars. In the latter, Steem market cap is essentially collateral; Steem Dollars can be redeemed for \$1 worth of newly minted Steem, and so redemptions affect all Steem hodlers via inflation. Since then, many new stablecoins have arisen based on similar ideas by UMA, Reflexer, and Liquity, as well as \emph{endogenous collateral} stablecoins like Synthetix sUSD, Terra UST, and Celo Dollar (see \cite{klagesmundt2020stablecoins} for further discussion). Notably, unlike custodial stablecoins, Dai is not currently considered as emoney or payment method subject to the Payment Services Directive in the European Union since there is no single issuer or custodian. Thus it does not have AML/KYC requirements.

In an academic white paper, \cite{cao2018} proposed a variation on cryptocurrency-collateralized DStablecoin design. It standardizes the speculative positions by restricting leverage to pre-defined bounds using automated resets. A consequence of these leverage resets is that stablecoin holders are partially liquidated from their positions during downward resets--i.e., when leverage rises above the allowed band due to  a cryptocurrency price crash. This compares with Dai, in which stablecoin holders are only liquidated in global settlement. An effect of this difference is that, in order to maintain a stablecoin position in the short-term, stablecoin holders need to re-buy into stablecoins (at a possibly inflated price) after downward resets. Of the many designs, it is unclear which deleveraging method would lead to a system that survives longer. This motivates us to study the dynamics of DStablecoin systems.

\begin{figure}
	\centering
	\begin{subfigure}[b]{0.4\textwidth}
		\includegraphics[width=\textwidth]{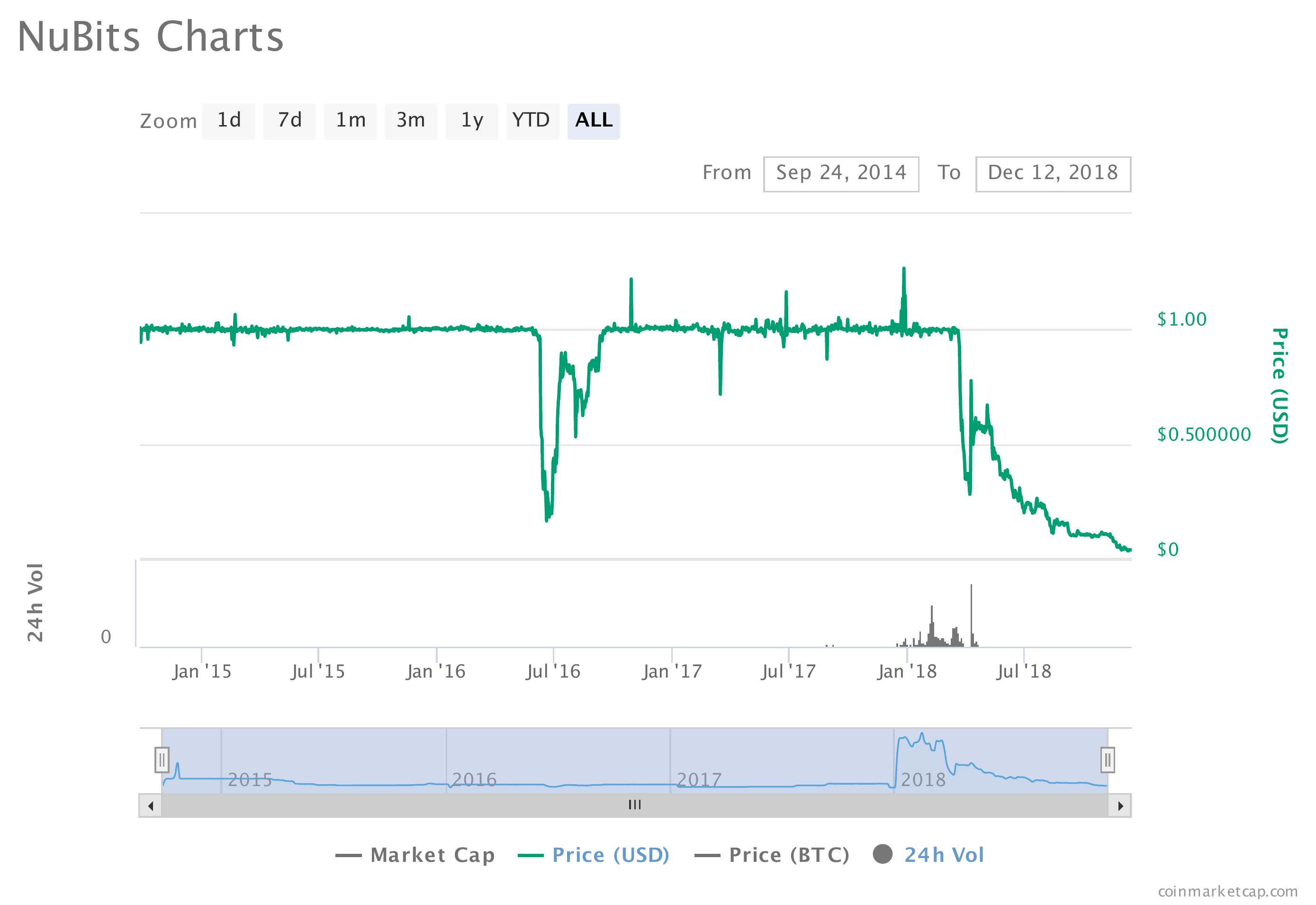}
		\caption{NuBits trades at cents on the dollar.}\label{fig:nubits_chart}
	\end{subfigure}
	\begin{subfigure}[b]{0.4\textwidth}
		\includegraphics[width=\textwidth]{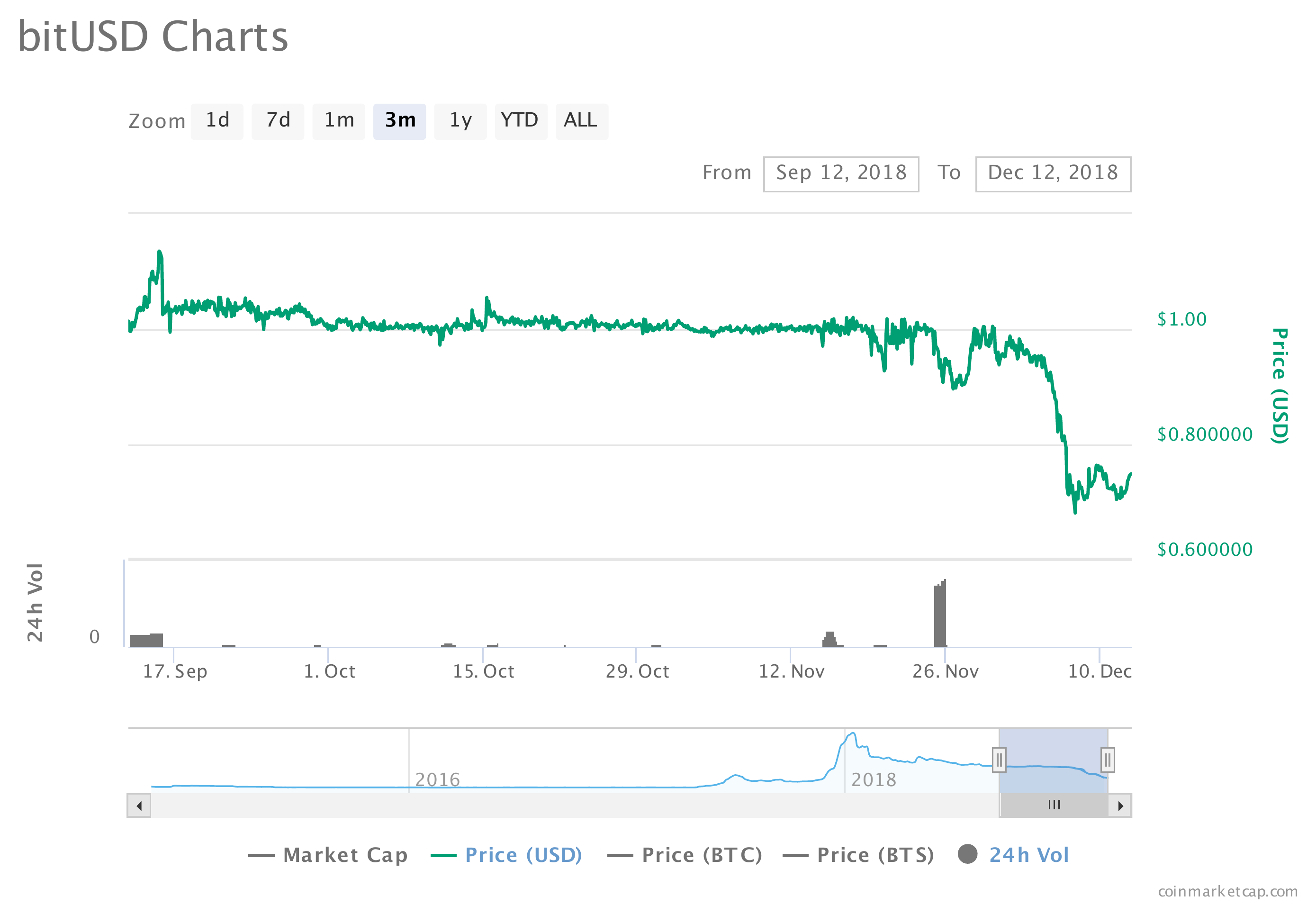}
		\caption{BitUSD has broken its USD peg.}\label{fig:bitusd_chart}
	\end{subfigure}
	\caption{Depeggings in decentralized stablecoins.}
\end{figure}

Non-custodial stablecoins have now experienced a wide array of volatility events, failures, and attacks.
Since the initial release of this paper in 2019, Black Thursday in March 2020 saw massive liquidation events result in a substantial depegging in Dai \cite{maker_spiral2020}, mirroring our results in Sections~\ref{sec:dynamics}-\ref{sec:stable_v_unstable}, and miner mempool manipulation that contributed to Dai liquidation auctions clearing at near zero prices at a cost of \$8m to the Maker system \cite{blocknative2020}, mirroring attack surfaces we described in Section~\ref{sec:attacks}.
Prior to this, as discussed in \cite{klagesmundt2018state}, \href{https://nubits.com/}{Nubits} has traded at cents on the dollar since 2018 (Figure~\ref{fig:nubits_chart}), and bitUSD and Steem Dollars have broken their USD pegs periodically (Figure~\ref{fig:bitusd_chart}).
Many additional examples of stablecoin mechanism failures and exploitations occurred through the rest of 2020 (see \cite{klagesmundt2020stablecoins,werner2021sok}).
Yet, the stablecoin space has remained heated with projects such as Dai growing rapidly and many new contenders arising, including UMA, Reflexer, Celo, and Liquity. The work in this paper has proven consequential for the progression of these projects (e.g., \cite{theblock_Maker2019,awesomeMakerDAO}).

\subsection{Relation to prior work}

Stablecoins are active cryptocurrencies, for which pre-existing models do not understand how the collateral rule enforces stability and how the interaction of different agents can affect stability.

With the notable exception of \cite{cao2018}, rigorous mathematical work on non-custodial stablecoins is lacking. They applied option pricing theory to valuing tranches in their proposed DStablecoin design using advanced PDE methods. In doing so, they need the simplifying assumption that DStablecoin payouts (e.g., from interest/fee payments and liquidations from leverage resets) are exogenously stable with respect to USD. This may circularly cause stability. In reality, these payouts are made in volatile cryptocurrency (ETH). From these ETH payments, stablecoin holders can
\begin{enumerate}
	\item Hold ETH and so take on ETH exposure,
	\item Use the ETH to re-buy into stablecoin, likely at an inflated price as it endogenously increases demand after a supply contraction,
	\item Convert the ETH to fiat, which requires waiting for block confirmations in an exchange (possibly hours) during times when ETH is particularly volatile and paying costs for fiat conversion (fees, potentially taxes). Notably, this is not available in all jurisdictions.
\end{enumerate}
To maintain a DStablecoin position, stablecoin holders need to re-buy into DStablecoins at each reset at endogenously higher price. Stablecoin holders additionally face the risk that the size of the DStablecoin market collapses such that the position cannot be maintained (and so ends up holding ETH). As no stable asset models exist to understand these endogenous effects, the analysis can't be easily extended using the traditional financial literature.\footnote{A secondary issue with their continuous model is that these systems are inherently discontinuous due to the discrete nature of incorporating blockchain transactions into blocks. Thus resets can occur beyond the set thresholds.} Our focus in this paper is complementary to understand these endogenous stable asset effects.

\cite{lipton2018} studied the evolution of custodial stablecoins. 
A few works on stablecoins have also arisen since the initial release of our paper.
\cite{klagesmundt2019vuln} described governance attack surfaces in non-custodial stablecoins, which is extended with general models in \cite{klagesmundt2020stablecoins}. \cite{evans2019ratings} presented an analysis of credit risk stemming from collateral type in the Maker system. And \cite{terra2, celo} modeled stability in the Terra and Celo stablecoins under different scenarios of Brownian motion without the endogenous market feedback effects we study in this paper.

In the context of central counterparty clearinghouses, the default fund contributions, margin requirements and participation incentives have been studied in, e.g., \cite{capponi2017}, \cite{amini2015}, and \cite{duffie2015}. The critical question in this area is understanding the effects of a liquidation policy of a member's portfolio in the case of a significant event. The counterpart of this in a decentralized setting is understanding the impact of DStablecoin deleveraging on system stability.

Stablecoin holders bear some resemblance to agents in currency peg and international finance models, e.g., \cite{morris1998} and \cite{guimaraes2007}. In these models, the market maker is essentially the government but is modeled with mechanical behavior and is not a player in the game. For instance, in \cite{guimaraes2007}, devaluation is modeled by a simple exogenous threshold rule: the government abandons the peg if the net demand for currency breaches the threshold and is otherwise committed to maintaining the peg. In contrast to currency markets, no agents are committed to maintaining the peg in DStablecoin markets. The best we can hope is that the protocol is well-designed and that the peg is maintained with high probability through the protocol's incentives. The role of government is replaced by decentralized speculators, who issue and withdraw stablecoins in a way to optimize profit. A fully strategic model would be a complicated dynamic game--these tend to be intractable and, indeed, are avoided in the currency peg literature in favor of a sequence of one period games. We enable a more endogenous modeling of speculators' optimization problems under a variety of risk constraints. Our model is a sequence of one-period optimization problems, in which dynamic coupling comes through the risk constraints.

DStablecoin speculators are similar to market makers in market microstructure models (e.g., \cite{ohara1997}). Like classical market microstructure, we do have a multi-period system with multiple agents subject to leverage constraints that take recurring actions according to their objectives. In contrast, in the DStablecoin setting, we do not have a truly stable asset that is efficiently and instantaneously available. Instead, agents make decisions that endogenously affect the price of the `stable' asset and affect the agents' future decisions and incentives to participate in a non-stationary way. In turn, the (in)stability results from the dynamics of these decisions.

Since the initial release of our paper in June 2019, \cite{klagesmundt2020} has described a complementary model of non-custodial stablecoins related to the model in this paper. That paper explores a different model of liquidation structure that affects speculator decision-making and applies martingale methods to analytically characterize stability. In contrast, in this paper we derive stability results about a simpler model that is more amenable to simulations, which we perform, and demonstrate stablecoin attacks that can arise from profitable bets against other agents.

\subsection{This paper}
We develop a dynamic model for non-custodial stablecoins that is complex enough to take into account the feedback effects discussed above and yet remains tractable. Our model can be interpreted as a market microstructure model in this new type of asset market.

Our model involves agents with different risk profiles; some desire to hold stablecoins and others speculate on the market. These agents solve optimization problems consistent with a wide array of documented market behaviors and well-defined financial objectives. As is common in the literature on market microstructure and currency peg games, these agents' objectives are myopic. These objectives are coupled for non-myopic risk using a flexible class of rules that are widely established in financial markets; these allow us to model the effects of a range of cyclic and counter-cyclic behaviors. The exact form of these rules is selected and self-imposed by speculators to match their desired responses and not part of the stablecoin protocol. Thus well-established manipulation of similar rules as applied to traditional financial regulation is not a problem here. Our model goes largely beyond a one-period model. We introduce this model with supporting rationale for design choices in Section~\ref{sec:model}.

Using our model, we make the following contributions:
\begin{itemize}
	\item We derive fundamental results about dynamics and liquidity in our model (Section\ref{sec:dynamics}).
	
	\item We demonstrate that stablecoins face deleveraging feedback effects that may cause illiquidity during crises and exacerbate collateral drawdown (Section~\ref{sec:deleveraging}).
	
	\item We characterize stable dynamics of the system under certain conditions that guarantee no liquidity crash (Section~\ref{sec:stable_v_unstable}) and show instability can occur in simulations outside of this setting (Section~\ref{sec:instability}).
	
	\item We simulate a wide range of market behaviors and find that speculator behavior has a large effect on realized volatilities, but that stablecoin failure times are largely determined by underlying asset movements (Section~\ref{sec:simulations}).
	
	\item We describe new attacks that exploit arbitrage-like opportunities around stablecoin liquidations (Section~\ref{sec:attacks}).
\end{itemize}
We relate these results to historical stablecoin events and apply these insights to suggest design improvements that aim to improve long-term stability. Based on these insights, we also suggest that interactions between multiple speculators and attackers may be the most interesting relationships to explore in more complex models.

\section{Model}\label{sec:model}

Our model couples a number of variables of interest in a risk transfer market between stablecoin holders and speculators. The stablecoin protocol dictates the logic of how agents can interact with the smart contracts that form the system; the design of this influences how the market plays out. Many DStablecoin designs have been proposed. We set up our model to emulate a DStablecoin protocol like Dai with global settlement, but the model is adaptable to different design choices. Note that our model is formulated with very few parameters given the problem complexity.

Our model builds on the model of traditional financial markets in \cite{farmer2015} but is new in design by incorporating endogenous stablecoin structure. In the model, we assume that the underlying consensus layer (e.g., blockchain) works well to confirm transactions without censorship or attack and that the system of contracts executes as intended.

\paragraph{Agents.} Two agents participate in the market.
\begin{itemize}
	\item The \textbf{stablecoin holder} seeks stability and chooses a portfolio to achieve this.
	\item The \textbf{speculator} chooses leverage in a speculative position behind the DStablecoin.
\end{itemize}

Stablecoin holders are motivated by risk aversion, trade limitations, and budget constraints. They are inherently willing to hold cryptoassets. In the current setting, this means they are likely either traders looking for short-term stability, users from countries with unstable fiat currencies, or users who are using cryptocurrencies to move money across borders. In the future, cryptocurrencies may be more accepted in economic exchange. In this case, stablecoin holders may be ordinary consumers who face risk aversion and budgeting for required consumption.

Speculators are motivated by (1) access to leverage and (2) security lending to borrow against their Ether holdings without triggering tax incidence or giving up Ether ownership. In order to begin participating, speculators need to either have confidence in the future of cryptocurrencies, think they can make money trading the markets, or face unusually high tax rates (or other barriers) that make security lending cheaper than outright selling assets. The model in this paper focuses on the first motivation. We propose an extension to the model that considers the second motivation.

\paragraph{Assets.} There are two assets. For simplicity, we give these assets specific names; however, they could be abstracted to other cryptocurrencies or outside of a cryptocurrency setting.
\begin{itemize}
	\item \textbf{Ether}:  high risk asset whose USD market prices $p^E_t$ are exogenous
	\item \textbf{DStablecoin}: a `stable' asset collateralized in Ether whose USD price $p^D_t$ is endogenous
\end{itemize}

Notably, a large DStablecoin system may have endogenous amplification effects on Ether price, similarly to how CDOs affected underlying assets in the 2008 financial crisis. We discuss this further in Section~\ref{sec:discussion} but leave formal modeling of this to future work.

There are several barriers for trading between crypto and fiat, which motivate our choice of assets. Most crypto-fiat pairs are through Bitcoin or Ether, which act as a gateway to other cryptoassets. Trading to fiat can involve moving assets between a number of exchanges and can take considerable time to confirm on the blockchain. Trading to a stablecoin is comparatively simple. Trading to fiat can also trigger more clear tax incidence. Additionally, some countries have imposed strict capital controls on trading between fiat and crypto.

\paragraph{Model outline.}
At $t=0$, the agents have endowments and prior beliefs. In each period $t$:
\begin{enumerate}
	\item New Ether price is revealed
	\item Ether expectations are updated
	\item Stablecoin holder decides portfolio weights
	\item Speculator, seeing demand, decides leverage
	\item DStablecoin market is cleared
\end{enumerate}

\subsection{Stablecoin holder}
The stablecoin holder starts with an initial endowment and decides portfolio weights to attain the desired stability. The following table defines the agent's state variables.
\begin{center}
	\begin{tabular}{c|l}
		\textbf{Variable}	&	\textbf{Definition} \\
		\hline
		$\bar n_t$		&	Ether held at time $t$ \\
		$\bar m_t$		&	DStablecoin held at time $t$ \\
		$\mathbf{w_t}$			&	Portfolio weights chosen at time $t$
	\end{tabular}
\end{center}

The stablecoin holder weights its portfolio by $\mathbf{w_t}$. We denote the components as $w^E_t$ and $w^D_t$ for Ether and DStablecoin weights respectively. The stablecoin holder's portfolio value at time $t$ is
$$\mathcal{A}_t = \bar n_t p^E_t + \bar m_t p^D_t = \bar n_{t-1} p^E_t + \bar m_{t-1} p^D_t.$$
Given weights, $\bar n_t$ and $\bar m_t$ will be determined based on the stablecoin clearing price $p_t^D$.

The basic results in Section~\ref{sec:dynamics} hold generally for any $\mathbf{w_t}\geq 0$ (i.e., there is no shorting). In this case, $\mathbf{w_t}$ could be chosen, e.g., from Sharpe ratio optimization, mean-variance optimization, or Kelly criterion (among others). In Sections~\ref{sec:stable_v_unstable}~\&~\ref{sec:simulations}, in order to focus on the effects of speculator decisions, we simplify the stablecoin holder as exogenous with unit price-elastic demand. In this case, DStablecoin demand is constant in dollar terms.

\subsection{Speculator}
The speculator starts with an endowment of Ether and initial beliefs about Ether's returns and variance and decides leverage to maximize expected returns subject to protocol and self-imposed constraints. The following tables define variables and parameters for the speculator.
\begin{center}
	\begin{tabular}{c|l}
		\textbf{Variable}	&	\textbf{Definition} \\
		\hline
		$n_t$		&	Ether held at time $t$ \\
		$r_t$		&	Expected return of Ether at time $t$ \\
		$\sigma^2_t$	&	Expected variance of Ether at time $t$ \\
		$\mathcal{L}_t$		&	Total stablecoins issued at time $t$ \\
		$\Delta_t$	&	Change to stablecoin supply at time $t$ \\
		$\tilde\lambda_t$	&	Leverage bound at time $t$
	\end{tabular}
\end{center}
\begin{center}
	\begin{tabular}{c|l}
		\textbf{Parameter}	&	\textbf{Definition} \\
		\hline
		$\gamma$	&	Memory parameter for return estimation \\
		$\delta$	&	Memory parameter for variance estimation \\
		$\beta$		&	Collateral liquidation threshold \\
		$\alpha$	&	Parameter governing risk measure (inversely related to VaR) \\
		$b$			&	Cyclicality parameter in risk constraint: pro- ($b>0$) or counter-cyclic ($b<0$)
	\end{tabular}
\end{center}

\subsubsection{Ether expectations} The speculator updates expected returns $r_t$, log-returns $\mu_t$ (used for the variance estimation), and variance $\sigma_t^2$ based on observed Ether returns as follows:
\begin{equation}
\begin{aligned}
r_t &= (1-\gamma) r_{t-1} + \gamma \frac{p^E_t}{p^E_{t-1}}, \\
\mu_t &= (1-\delta)\mu_{t-1} + \delta \log \frac{p_t^E}{p^E_{t-1}}, \\
\sigma_t^2 &= (1-\delta) \sigma_{t-1}^2 + \delta \Big( \log \frac{p^E_t}{p^E_{t-1}} - \mu_t\Big)^2.
\end{aligned}
\end{equation}\label{eq:expectations}
For fixed memory parameters $\gamma,\delta$ (lower memory parameter = longer memory), these are exponential moving averages consistent with the RiskMetrics approach commonly used in finance \cite{longerstaey1996}. For sufficiently stepwise decreasing memory levels and assuming i.i.d. returns, this process will converge to the true values supposing they are well-defined and finite. In reality, speculators don't outright know the Ether return distribution and, as we will see in the simulations, the stablecoin system dynamics occur on timescales shorter than required for convergence of expectations. Thus, we focus on the simpler case of fixed memory parameters.

Note that $\gamma \neq \delta$ may be reasonable. Current cryptocurrency markets are not very price efficient, and so traders might reasonably take into account momentum when estimating returns while using a wider memory for estimating covariance.

We additionally consider the case in which the speculator knows the Ether distribution outright and $\gamma=\delta=0$. This is consistent with a rational expectations standpoint but ignores how the speculator arrives at that knowledge.

\subsubsection{Optimize leverage: choose $\Delta_t$}

The speculator is liable for $\mathcal{L}_t$ DStablecoins at time $t$. At each time $t$, it decides the number of DStablecoins to create or repurchase. This changes the stablecoin supply $\mathcal{L}_t = \mathcal{L}_{t-1} + \Delta_t$. If $\Delta_t>0$, the speculator creates and sells new DStablecoin in exchange for Ether at the clearing price. If $\Delta_t<0$, the speculator repurchases DStablecoin at the clearing price.

Strictly speaking, the speculator will want to maximize its long-term withdrawable value. At time $t$, the speculator's withdrawable value is the value of its ETH holdings minus collateral required for any issued stablecoins: $n_t p_t^E - \beta\mathcal{L}_t$. Maximizing this is not amenable to a myopic view, however, as maximizing the next step's withdrawable value is only a good choice when the speculator intends to exit in the next step.

Instead, we frame the speculator's objective as maximizing expected equity: $n_t p_t^E - \mathbf{E}[p^D] \mathcal{L}_t$. In this, the speculator expects to be able to settle liabilities at a long-term expected value of $\mathbf{E}[p^D]$. The market price of DStablecoin will fluctuate above and below \$1 naturally depending on prevailing market conditions. The actual expected value is nontrivial to compute as it depends on the stability of the DStablecoin system. For individual speculators with small market power, we argue that $\mathbf{E}[p^D]=1$ is a an assumption they may realistically make, as we discuss further below. This is additionally the value realized in the event of global settlement.

We suggest that this optimization is a candidate for `honest' behavior of a speculator as it is consistent with the speculator acting on perceived arbitrage in mispricings of DStablecoin from the peg. In essence, the speculator expects to increase (reduce) leverage `at a discount' when $p_t^D$ is above (below) target. This is the typically cited mechanism by which these systems maintain their peg and thus how the designers \emph{intend} for speculators to behave. However, this assumes that $p_t^D$ is sufficiently stable/mean-reverting to \$1 and so this behavior may not in fact be a best response.

\paragraph{Aggregate vs. individual speculators.}
In our model, the single speculative agent, which is not a price-taker, is intended to reflect the aggregate behavior of many individual speculators, each with small market power.\footnote{We propose to relax this simplification in follow-up work by considering the interaction of many speculators with longer term strategic thinking.} In a normal liquid market, an individual speculator would be able to repurchase DStablecoins at dollar cost and walk away with the equity. By maximizing equity, the aggregate speculator considers its liabilities to be \$1 per DStablecoin. This may turn out to be untrue during liquidity crises as the repurchase price may be higher. In our model, speculator's don't know the probability of crises and instead account for this in a conservative risk constraint.

\paragraph{Formal optimization problem.}
The speculator chooses $\Delta_t$ by maximizing expected equity in the next period subject to a leverage constraint:
$$\begin{aligned}
\max_{\Delta_t} \hspace{0.5cm} & r_t\Big(n_{t-1}p^E_t + \Delta_t p^D_t(\mathcal{L}_t)\Big) - \mathcal{L}_t \\
\text{s.t.} \hspace{0.5cm} & \Delta_t \in \mathcal{F}_t
\end{aligned}$$
where $\mathcal{F}_t$ is the feasible set for the leverage constraint. This is composed of two separate constraints: (1) a \textbf{liquidation constraint} that is fundamental to the protocol, and (2) a \textbf{risk constraint} that encodes the speculator's desired behavior. Both are introduced below.

If the leverage constraint is unachievable, we assume the speculator enters a `recovery mode', in which it tries to maximize its chances of returning to the normal setting. In this case, it solves the optimization using only the liquidation constraint. If the liquidation constraint is unachievable, the DStablecoin system fails with a global settlement.

\subsubsection{Liquidation constraint: enforced by the protocol}
The liquidation constraint is fundamental to the DStablecoin protocol. A speculator's position undergoes forced liquidation at time $t$ if either (1)~after $p_t^E$ is revealed, $n_{t-1} p^E_t < \beta \mathcal{L}_{t-1}$, or (2)~after $\Delta_t$ is executed, $n_t p_t^E < \beta \mathcal{L}_t$. The speculator aims to control against this as liquidations can occur at unfavorable prices and are associated with fees in existing protocols (we exclude these fees from our simple model, but they can be easily added).

Define the speculator's leverage as the $\beta$-weighted ratio of liabilities to assets\footnote{We choose this definition to simplify the model. The alternative definition $\lambda'~=~\frac{\text{assets}}{\text{assets} - \beta\cdot\text{liabilities}}$ describes the same idea scaled from 0 to $\infty$. I.e., $\lambda' = \frac{1}{1-\lambda}$ is monotonically increasing in $\lambda$ for $0\leq \lambda' < 1$.}

$$\lambda_t = \frac{\beta \cdot \text{liabilities}}{\text{assets}}.$$
The liquidation constraint is then $\lambda_t \leq 1$.

\subsubsection{Risk constraint: self-imposed speculator behavior}
The risk constraint encodes the speculator's desired behavior into the model. \emph{We assume no specific type for the risk constraint in our analytical results, which are generic.} For our simulations, we explore a variety of speculator behaviors via the risk constraint. We first consider Value-at-Risk (VaR) \emph{as an example} of a constraint realistically used in markets. This is consistent with narratives shared by Dai speculators about leaving a margin of safety to avoid liquidations. We then construct a generalization that goes well beyond VaR and allows us to explore a spectrum of pro-cyclical and counter-cyclical behaviors encoded in the risk constraint.

Manipulation and instability resulting from similar \emph{externally-imposed} VaR rules is a well-known problem in the risk management and financial regulatory literature (see e.g., \cite{farmer2015}). This is of less concern here as the precise parameters of the risk constraint are selected and self-imposed by speculators to approximate their own utility optimization and are not part of the DStablecoin protocol. Further, we consider constraints that go \emph{beyond VaR}. We instead need to show that our results are robust to a variety of risk constraints that speculators could select.

\paragraph{Example: VaR-based constraint.}
The VaR-based version of the risk constraint is
$$\lambda_t \leq \exp(\mu_t - \alpha \sigma_t),$$
where $\alpha>0$ is inversely related to riskiness. This is consistent with VaR for normal and maximally heavy-tailed symmetric return distributions with finite variance.

Let $\text{VaR}_{a,t}$ be the $a$-quantile per-dollar VaR of the speculator's holdings at time $t$. This is the minimum loss on a dollar in an $a$-quantile event. With a VaR constraint, the speculator aims to avoid triggering the liquidation constraint in the next period with probability $1-a$, i.e.,
$\mathbf{P} \Big( n_t p^E_{t+1} \geq \beta \mathcal{L}_t \Big) \geq 1-a.$ To achieve this, the speculator chooses $\Delta_t$ such that
$$\Big(n_{t-1}p^E_t + \Delta_t p_t^D(\mathcal{L}_t)\Big) (1-\text{VaR}_{a,t}) \geq \beta \mathcal{L}_t.$$
This requires $\lambda_t \leq 1-\text{VaR}_{a,t}$, which addresses the probability that the liquidation constraint is satisfied next period and implies that it is satisfied this period.

Define $\tilde \lambda_t := \exp(\mu_t -\alpha\sigma_t)$. Then $\tilde \lambda_t$ is increasing in $\mu_t$ and decreasing in $\sigma_t$. Further, the fatter the speculator thinks the tails of the return distribution are, the greater $\alpha$ will be, and the lesser $\tilde \lambda_t$ will be, as we demonstrate next.

\paragraph{VaR constraint with normal returns.}
If the speculator assumes Ether log returns are $(\mu_t, \sigma_t)$ normal, then
$\text{VaR}_{a,t} = 1 - \exp\Big(\mu_t + \sqrt{2} \sigma_t \text{erf}^{-1}(2a-1)\Big).$
Defining $\alpha = - \sqrt{2}\text{erf}^{-1}(2a-1)$, which is positive for appropriately small $a$, the VaR constraint is
$\lambda_t \leq 1 - \text{VaR}_{a,t} = \exp(\mu_t - \alpha \sigma_t).$

\paragraph{VaR constraint with heavy tails.}
If Ether log returns $X$ are symmetrically distributed with finite mean $\mu_t$ and finite variance $\sigma_t^2$, then for any $\alpha>1$, Chebyshev's inequality gives us
$$\mathbf{P}(X < \mu_t -\alpha\sigma_t) \leq \frac{1}{2\alpha^2}.$$
For the maximally heavy-tailed case, this inequality is tight. Then for VaR quantile $a$, we can find the corresponding $\alpha$ such that $a = \frac{1}{2\alpha^2}$. The log return VaR is $\mu_t-\alpha\sigma_t$, which gives the per-dollar $\text{VaR}_{a,t} = 1-\exp(\mu_t-\alpha\sigma_t)$. Then the VaR constraint is $\lambda_t \leq \exp(\mu_t - \alpha\sigma_t)$.

\paragraph{Generalized risk constraint.}
Similarly to \cite{farmer2015}, we can generalize the bound to explore a spectrum of different behaviors:
$$\ln \tilde \lambda = \mu_t - \alpha \sigma_t^b,$$
where $\alpha$ is an inverse measure of riskiness and $b$ is a cyclicality parameter. A positive $b$ means that $\tilde \lambda_t$ decreases with perceived risk (pro-cyclical). A negative $b$ means that $\tilde \lambda_t$ increases with perceived risk (counter-cyclical). 

\subsection{DStablecoin market clearing}

The DStablecoin market clears by setting demand = supply in dollar terms:
$$w^D_t \Big(\bar n_{t-1}p^E_t + \bar m_{t-1}p^D_t(\mathcal{L}_t)\Big) = \mathcal{L}_t p^D_t(\mathcal{L}_t).$$
The demand (left-hand side) comes from the stablecoin holder's portfolio weight and asset value. Notice that while the asset value depends on $p_t^D$, the portfolio weight $w_t^D$ does not. That is, the stablecoin holder buys with market orders based on weight. This simplification allows for a tractable market clearing; however, it is not a full equilibrium model.

We justify this choice of simplified market clearing with the following observations:
\begin{itemize}
	\item The clearing is similar to constant product market maker model used in the Uniswap decentralized exchange (DEX) \cite{zhang2018}.
	\item Sophisticated agents are known to be able to front-run DEX transactions \cite{daian2019}. As speculators are likely more sophisticated than ordinary stablecoin holders, in many circumstances they can see demand before making supply decisions.\footnote{This said, DEX mechanics differ slightly from our specific formulation. To make the model more realistic, stablecoin holders could issue buy offers in token units instead of weights at the expense of greater model complexity.}
	\item Evidence from Steem Dollars suggests that demand need not decrease tremendously with price in the unique setting in which stable assets are not efficiently available. Steem Dollars is a stablecoin with a mechanism for price `floor' but not `ceiling'. Over significant stretches of time, it has traded at premiums of up to $15\times$ target.
\end{itemize}

In most of our results, the time period context is clear. To simplify notation, in a given time $t$, we drop subscripts and write with the following quantities:

\begin{center}
	\begin{tabular}{l c l}
		\textbf{Quantity}					&\textbf{Sign}&\textbf{Interpretation} \\
		\hline
		$x:=w^D_t \bar n_{t-1} p^E_t$ 		&$x\geq 0$	& New DStablecoin demand available	\\
		$y:=w^D_t \bar m_{t-1} - \mathcal{L}_{t-1}$	&	$y\leq 0$	& $|y| = $ `free supply' in DStablecoin market \\
		$z:=n_{t-1} p^E_t$ 			&	$z\geq 0$	& Speculator value available to maintain market \\
		\hline
	\end{tabular}
\end{center}
\begin{center}
	\begin{tabular}{r c l}
		\hline
		$\mathcal{L}$	&$:=$&	$\mathcal{L}_{t-1}$ \\
		$\Delta$		&$:=$&	$\Delta_t$ \\
		$\tilde \lambda$&$:=$&	$\tilde \lambda_t$ \\
		$\mathbf{w}$	&$:=$&	$\mathbf{w_t}$ \\
		\hline
	\end{tabular}
\end{center}

With $\Delta > y$, which turns out to be always true as discussed later, the clearing price is
$$p_t^D(\Delta) = \frac{x}{\Delta-y}.$$

As the model is defined thus far, stablecoin holders only redeem coins for collateral through global settlement. However, this assumption is easily relaxed to accommodate algorithmic or manual settlements.

\section{Stable Asset Market Dynamics}\label{sec:dynamics}

We derive tractable solutions to the proposed interactions and results about liquidity and stability.

\subsection{Solution to the speculator's decision}

We first introduce some basic results about the speculator's leverage optimization problem.

\paragraph{Solving the leverage constraint.}
\begin{proposition}\label{prop:constraint_sol}
	Let $\Delta_{\min} \geq \Delta_{\max}$ be the roots of the polynomial in $\Delta$
	$$-\beta \Delta^2 + \Delta\Big( \tilde \lambda (z+x) - \beta(\mathcal{L} - y)\Big) - \tilde\lambda zy + \beta\mathcal{L} y.$$
	Assuming $\Delta > y$,
	\begin{itemize}
		\item If $\Delta_{\min},\Delta_{\max} \in \mathbb{R}$, then $[\Delta_{\min},\Delta_{\max}]\cap(y,\infty)$ is the feasible set for the leverage constraint.
		\item If the roots are not real, then the constraint is unachievable.
	\end{itemize}
\end{proposition}

\begin{center} \hyperlink{pf:constraint_sol}{\texttt{[Link to Proof]}} \end{center}

Setting $\tilde \lambda = 1$ gives the expression for the liquidation constraint alone.

The condition $\Delta > y$ makes sense for two reasons. First, if $\Delta < y$ then $p^D_t < 0$. Second, as we show below, the limit $\lim_{\Delta \rightarrow y^+} p^D_t = \infty$. Thus, if we start in the previous step under the condition $\Delta > y$, then the speculator will never be able to pierce this boundary in subsequent steps. We further discuss the implications of this condition later.

\paragraph{Solving the leverage optimization.}
\begin{proposition}\label{prop:leverage_sol}
	Assume that the speculator's constraint is feasible and let $[\Delta_{\min}, \Delta_{\max}] \cap (y, \infty)$ be the feasible region. Define $r:=r_t$, let $\Delta^* = y + \sqrt{-yrx}$, and define
	$$f(\Delta) = r\Delta \frac{x}{\Delta - y} - \Delta.$$
	Then the solution to the speculator's optimization problem is
	\begin{itemize}
		\item $\Delta^*$ if $\Delta^* \in [\Delta_{\min}, \Delta_{\max}] \cap (y, \infty)$
		\item $\Delta_{\min}$ if $\Delta^*<\Delta_{\min}$
		\item $\Delta_{\max}$ if $\Delta^*>\Delta_{\max}$
	\end{itemize}
\end{proposition}

\begin{center} \hyperlink{pf:leverage_sol}{\texttt{[Link to Proof]}} \end{center}

\subsection{Maintenance condition for the stable asset market}

The next result describes a bound to the speculator's ability to maintain the market. This bound takes the form of
\begin{center}
	(a lower bound on collateral) - (capital available to enter the market),
\end{center}
which must be sufficiently high for the system to be maintainable.
\begin{proposition}\label{prop:feasible_condition}
	The feasible set for the speculator's liquidation constraint is empty when
	$$\Big(\tilde \lambda(x+z) - \beta\mathcal{L} w^D \Big)^2 < 4\beta \tilde\lambda \mathcal{L}xw^E$$
\end{proposition}

\begin{center} \hyperlink{pf:feasible_condition}{\texttt{[Link to Proof]}} \end{center}

In Prop.~\ref{prop:feasible_condition}, $\beta\mathcal{L}w^D \geq 0$ is interpreted as a lower bound on the capital required to maintain the DStablecoin market into the next period (i.e., the collateral required for the minimum size of the DStablecoin market), $\tilde\lambda \in [0,1]$, and $x+z \geq 0$ is the capital available to enter the DStablecoin market from both the supply and demand sides. The inequality then states that the difference between the capital available to enter the market and the lower bound maintenance capital must be sufficiently high for the market to be maintainable by the speculator. The constraint $\Delta < y$ implies that the case of the negative difference does not work.

\subsection{Deleveraging effects, limits to market liquidity}\label{sec:deleveraging}

\paragraph{Limits to the speculator's ability to decrease leverage.}
The next result presents a fundamental limit to how quickly the speculator can reduce leverage by repurchasing DStablecoins, given the modeled market structure. Note that this limit applies even if the speculator can bring in additional capital. The term $-y = \mathcal{L}(1-w^D)$ represents the `free supply' of DStablecoin available for exchange, which can be increased by a positive $\Delta$.

\begin{proposition}\label{prop:liquidity_limit}
	The speculator with asset value $z$ cannot decrease DStablecoin supply at $t$ more than
	$$\Delta^- := \frac{z}{z+x}y.$$
	Further, even with additional capital, the speculator cannot decrease the DStablecoin supply at $t$ by more than $y$.
\end{proposition}

\begin{center} \hyperlink{pf:liquidity_limit}{\texttt{[Link to Proof]}} \end{center}

\paragraph{Deleveraging affects collateral drawdown through liquidity crises.}
The result leads to a DStablecoin market price effect from leverage reduction. This can lead to a \emph{deleveraging spiral}, which is a feedback loop in leverage reduction and drying liquidity. In this, the speculator repurchases DStablecoin to reduce leverage at increasing prices as liquidity dries up as repurchase tends to push up $p_t^D$ if outside demand remains the same. At higher prices, more collateral needs to be sold to achieve deleveraging, leaving relatively less in the system. Subsequent deleveraging, whether voluntary or through liquidation, becomes more difficult as the price effects compound.

Whether or not a spiraling effect occurs will depend on the demand behavior of stablecoin holders. The action of the stablecoin holder may actually exacerbate this effect: during extreme Ether price crashes, stablecoin holders will tend to increase their DStablecoin demand in a `flight to safety' move. Table~\ref{table:delev_spiral} illustrates an example scenario of a deleveraging spiral in a simplified setting with constant unit demand elasticity and in which the speculator's risk constraint is the liquidation constraint. Similar results hold under other constant demand elasticities. The system starts in a steady state. the Ether price declines trigger three waves of liquidations, forcing the speculator to liquidate her collateral to deleverage at rising costs.

\begin{table}
	\centering
	\begin{tabular}{c|c|c|c|c|c}
		$t$ & $p_t^E$ & $\Delta_t$ & $\mathcal{L}_t$ & $p_t^D$ & $n_t$ \\
		\hline
		$0$ & $85$ & & $100.583$ & $0.994$ & $1.8$ \\
		$1$ & $83$ & $-3.115$ & $97.468$ & $1.026$ & $1.761$ \\
		$2$ & $82$ & $-4.105$ & $93.363$ & $1.071$ & $1.708$ \\
		$3$ & $81$ & $-4.57$ & $88.793$ & $1.126$ & $1.644$ \\
	\end{tabular}
	\caption{Example scenario of a deleveraging spiral.}\label{table:delev_spiral}
\end{table}

If Ether prices continue to go down,\footnote{Ether price decline can further be facilitated by feedback from large liquidations, as discussed earlier.} the deleveraging spiral is only fixed if (1) more money comes into the collateral pool to create more DStablecoins, or (2) people lose faith in the system and no longer want to hold DStablecoins, which can cause the system to fail. There is no guarantee that (1) always happens.

This liquidity effect on DStablecoin price makes sense because the stablecoin (as long as it's working) should be worth more than the same dollar amount of ETH during a downturn because the stablecoin comes with additional protection. If the speculator is forced to buy back a sizeable amount of the coin supply, it will have to do so at a premium price.

One might think the spiral effect is good for stablecoin holders. As we explore in Section~\ref{sec:attacks}, this can be the case for a short-term trade. However, as we will see, the speculator's ability to maintain a stable system may deteriorate during these sort of events as it has less control or less willingness to control the coin supply. Deleveraging effects can siphon off collateral value, which can be detrimental to the system in the long-term.

This suggests the question: do alternative non-custodial designs suffer similar deleveraging problems? We compare to an alternative design described in \cite{cao2018}. In this design, the stablecoin is restricted to pre-defined leverage bounds, at which algorithmic `resets' partially liquidate both stablecoin holder and speculator positions at \$1 price. While this quells the price effect on collateral, it \emph{shifts} the deleveraging risk from speculator to stablecoin holder. The stablecoin holder is liquidated at \$1 price but, if they want to maintain a stablecoin position, they have to re-buy in to a smaller market at inflated price. Of the many designs, it is unclear which deleveraging method would lead to a system that survives longer.

\begin{figure}
	\centering
	\begin{subfigure}[b]{0.49\textwidth}
		\includegraphics[width=\textwidth]{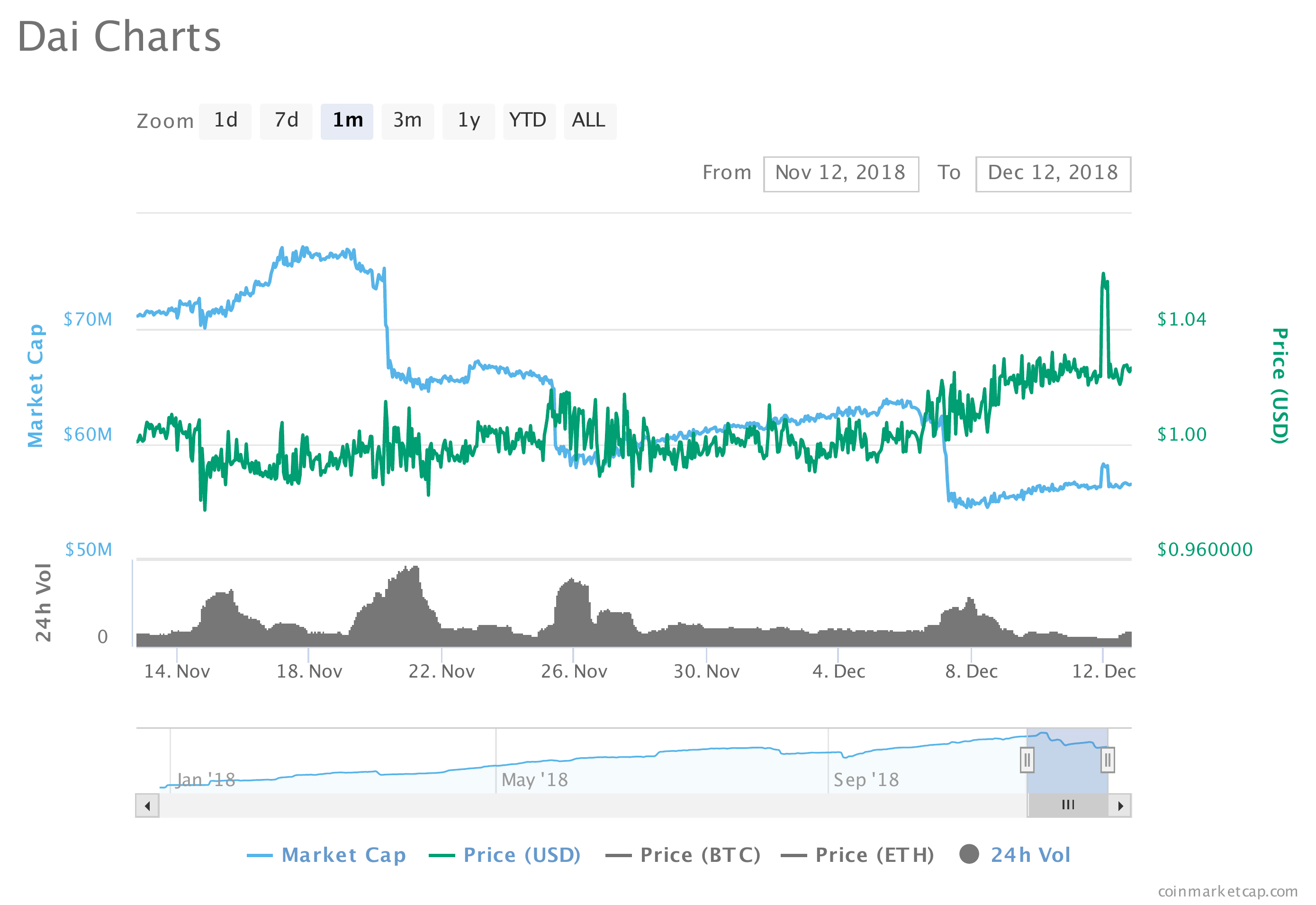}
		\caption{}\label{fig:dai_chart}
	\end{subfigure}
	\begin{subfigure}[b]{0.49\textwidth}
		\includegraphics[width=\textwidth]{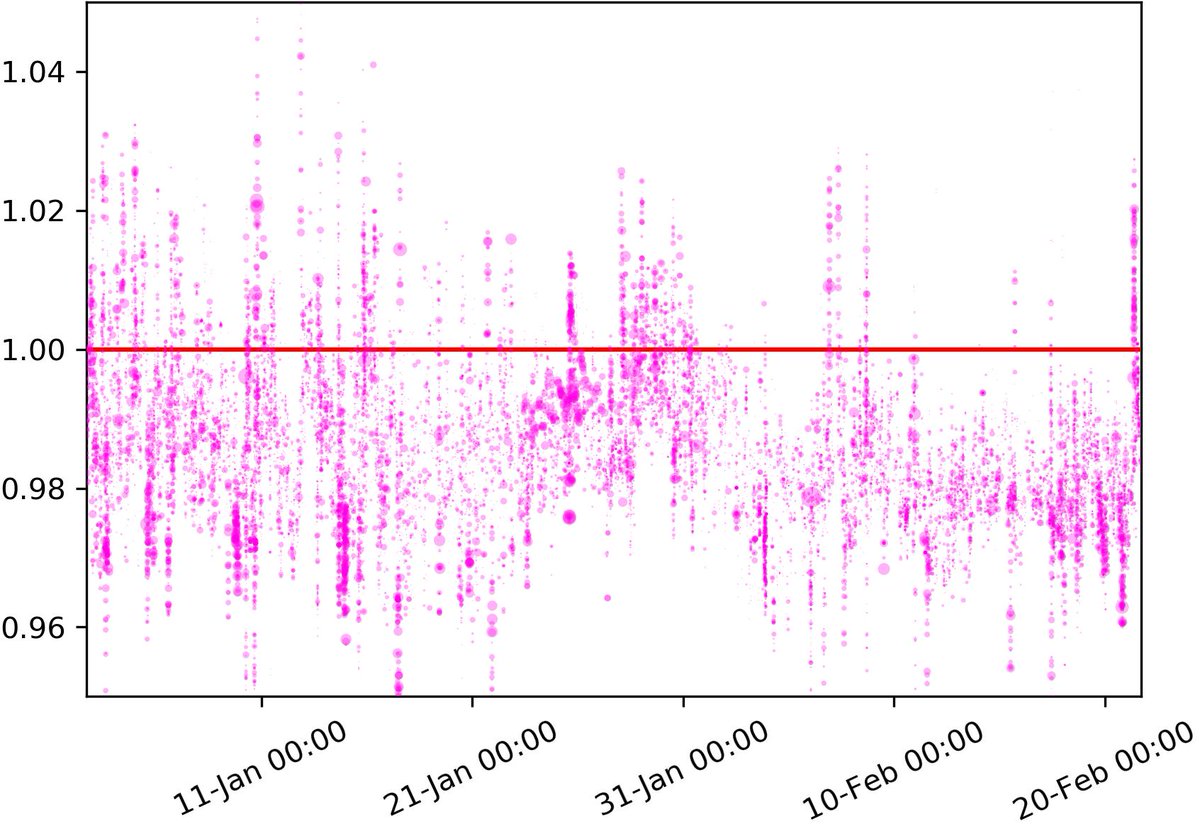}
		\caption{}\label{fig:dai_below_target}
	\end{subfigure}
	\caption{Model Results explain data from Dai market. (a) Dai deleveraging feedback in Nov-Dec 2018 (image from coinmarketcap). (b) Dai normally trades below target with spikes in price due to liquidations (image from dai.stablecoin.science).}\label{fig:real_liquidity}
\end{figure}

\paragraph{Results explain real market data.}
A preliminary analysis of Dai market data suggests that our results apply. Figure~\ref{fig:dai_chart} shows the Dai price appreciate in Nov-Dec 2018 during multiple large supply decreases. This is consistent with an early phase of a deleveraging spiral. Figure~\ref{fig:dai_below_target} shows trading data from multiple DEXs over Jan-Feb 2019: price spikes occur in the data reportedly from speculator liquidations \cite{rowe_tweet2019}. This provides empirical evidence that liquidity is indeed limited for lowering leverage in Dai markets. Further, as discussed in the next section, Dai empirically trades below target in many normal circumstances.

\begin{figure}
	\centering
	\begin{subfigure}[b]{0.49\textwidth}
		\includegraphics[width=\textwidth]{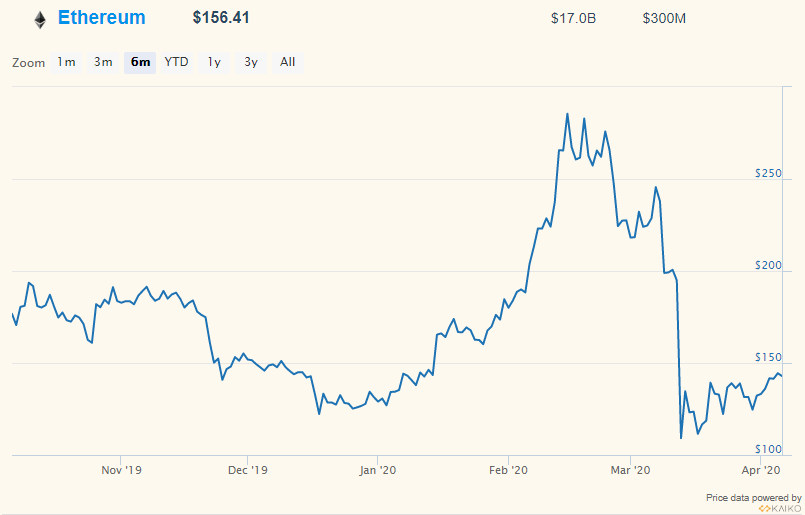}
		\caption{}\label{fig:eth_mar20}
	\end{subfigure}
	\begin{subfigure}[b]{0.49\textwidth}
		\includegraphics[width=\textwidth]{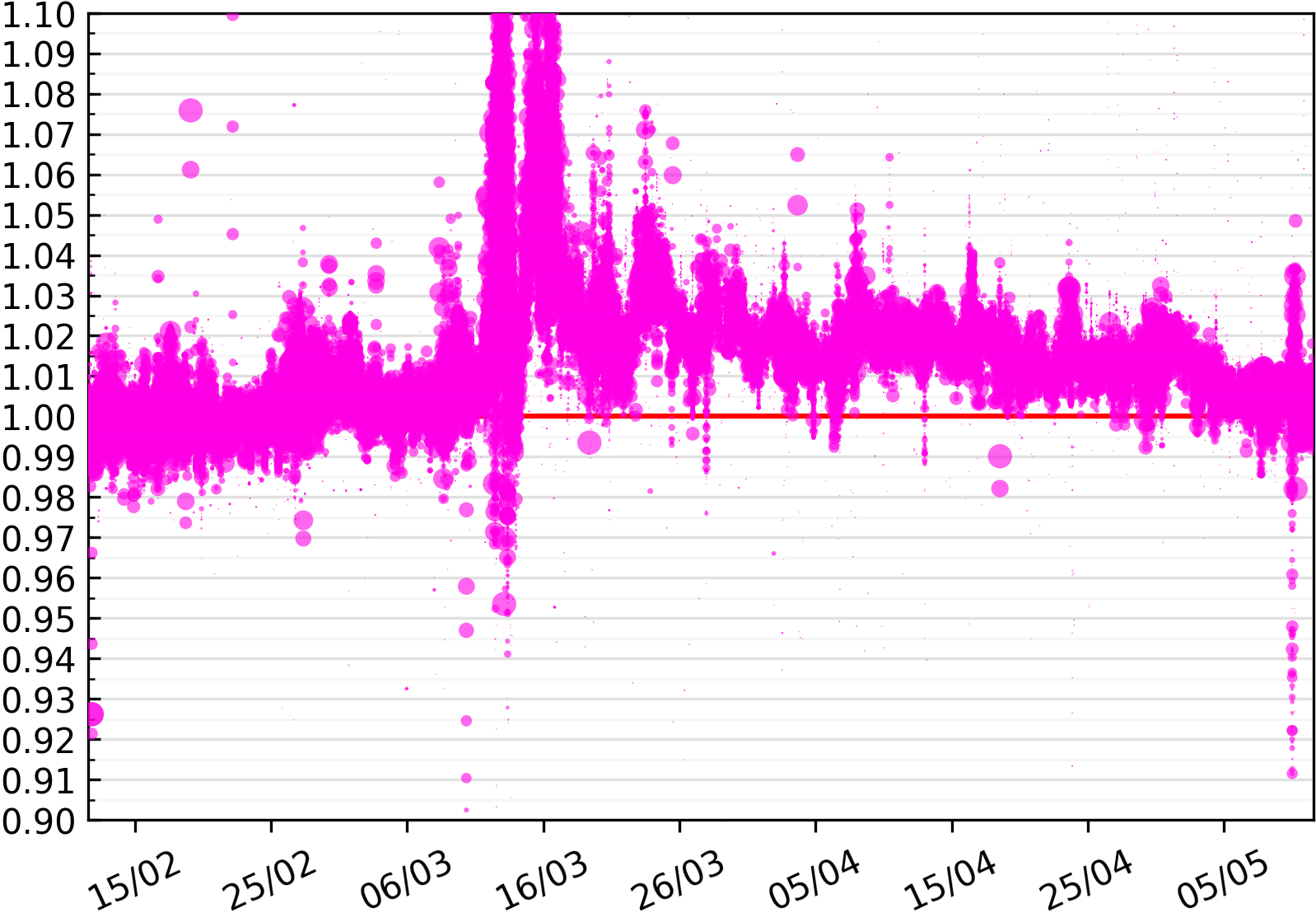}
		\caption{}\label{fig:dai_mar20}
	\end{subfigure}
	\caption{Black Thursday in March 2020. (a) $\sim 50\%$ ETH price crash (image from OnChainFX). (b) liquidation price effect on Dai DEX trades (image from dai.stablecoin.science).}\label{fig:mar20}
\end{figure}

Since releasing the initial version of this paper in June 2019, massive liquidation events around Black Thursday in March 2020 provide additional strong evidence of deleveraging effects in the Dai market. Figure~\ref{fig:eth_mar20} depicts a $\sim 50\%$ ETH price cash on 12 Mar. 2020, which precipitated a cascade of cryptocurrency liquidations. Figure~\ref{fig:dai_mar20} depicts the price effects of these liquidations on Dai prices on DEXs. Speculators deleveraging during this event had to pay premiums of $\sim 10\%$ and face consistent premiums $>2\%$ weeks into the aftermath. 
Concurrently, Maker was affected by global mempool flooding on Ethereum. This additionally contributed to Dai liquidation auctions clearing at near zero prices, which may in fact have amplified the deleveraging feedback effects. Altogether, Dai traded at significant premiums over this time despite Maker being in a much riskier state in terms of collateral and liquidations.
See \cite{klagesmundt2020insights} and \cite{blocknative2020} for further discussion of this event.

\section{Stability results}\label{sec:stable_v_unstable}

We now characterize stable price dynamics of DStablecoins when the leverage constraint is non-binding. For this section, we make the following simplifications to focus on speculator behavior:
\begin{itemize}
	\item The market has fixed dollar demand at each $t$: $w^D_t \mathcal{A}_t = \mathcal{D}$. This is consistent with the stablecoin holder having unit-elastic demand, or having an exogenous constraint to put a fixed amount of wealth in the stable asset.
	\item Speculator's expected Ether return is constant $r_t = \hat r>1$. This means they always want to fully participate in the market and is consistent with $\gamma=0$.
\end{itemize}
This amounts to setting $x = \mathcal{D}$ and $y=-\mathcal{L}$. Now the DStablecoin market clearing price is
$p^D_t = \frac{\mathcal{D}}{\mathcal{L}_t}.$
The leverage constraint (assuming $\mathcal{L} + \Delta > 0$) becomes
$$-\beta\Delta^2 + \Delta(\tilde\lambda(z+\mathcal{D}) - 2\beta\mathcal{L}) + \mathcal{L}(\tilde\lambda z - 2\beta - \beta\mathcal{L}) \geq 0.$$

The speculator's maximization objective becomes
$\hat r\Delta \frac{\mathcal{D}}{\mathcal{L}+\Delta} - \Delta,$
which gives
$$\Delta^* = -\mathcal{L} + \sqrt{\mathcal{L}\mathcal{D}\hat r}.$$

While we prove a stability result in this simplified setting, we believe the results can be extended beyond the assumption of constant unit-elastic demand.

\subsection{Stability if leverage constraint is non-binding}
\begin{proposition}\label{prop:stable1}
	Assume $w_t^D \mathcal{A}_t = \mathcal{D}$ (DStablecoin dollar demand) and $r_t = \hat r$ (speculator's expected Ether return) remain constant. If the leverage constraint is inactive at time $t$, then the DStablecoin return is
	$$\frac{p^D_t}{p^D_{t-1}} = \sqrt{\frac{\mathcal{L}}{\mathcal{D}\hat r}}.$$
\end{proposition}

\begin{center} \hyperlink{pf:stable1}{\texttt{[Link to Proof]}} \end{center}

Supposing that $\mathcal{D}\approx \mathcal{L}$ (i.e., the previous price was close to the \$1 target) and the constraint is inactive, Prop.~\ref{prop:stable1} tells us that the DStablecoin behaves stably like the payment of a coupon on a bond.

Consider estimators for DStablecoin log returns $\bar \mu_t$ and volatility $\bar \sigma_t$ computed in a similar way to Ether expectations in Eq.~\ref{eq:expectations}. When the leverage constraint is non-binding, DStablecoin log returns remain $\bar \mu_t \approx 0$, the contribution to volatility at time $t$ is $\ln \frac{p_t^D}{p_{t-1}^D} - \bar \mu_t \approx 0$, and the DStablecoin tends toward a steady state with stable price and zero variability. The next theorem formalizes this result to describe stable dynamics of price and the volatility estimator under the condition that the system doesn't breach the speculator's leverage threshold.

\begin{theorem}\label{prop:stable2}
	Assume $w_t^D \mathcal{A}_t = \mathcal{D}$ (DStablecoin demand) and $r_t = \hat r$ (speculator's expected Ether return) remain constant. Let $\mathcal{L}_0=\mathcal{D}$ and $\bar \mu_0, \bar \sigma_0$ be given. If the leverage constraint remains inactive through time $t$, then
	$$\mathcal{L}_t = \mathcal{D}\hat{r}^{\frac{2^t-1}{2^t}},
	\hspace{1cm} \bar \mu_t = \begin{cases}
	(1-\delta)^t \bar \mu_0 - \delta \frac{(1-\delta)^t-2^{-t}}{2(1-\delta)-1} \ln \hat r, & \text{ if } \delta \neq 1/2 \\
	2^{-t}\Big( \bar \mu_0 - \frac{1}{2} t \ln \hat r \Big), & \text{ if } \delta = 1/2
	\end{cases}$$
	$$\bar\sigma_t^2 = \begin{cases}
	\sum_{k=1}^t (1-\delta)^{t-k}\delta \Big( (1-\delta)^k \bar \mu_0 - \frac{(1-\delta)^k -2^{-k+1}(1-\delta)}{2(1-\delta)-1}\ln \hat r \Big)^2 + (1-\delta)^t\bar\sigma_0^2, & \text{ if } \delta \neq 1/2 \\
	2^{-t} \sum_{k=1}^t 2^{-k-1} \Big( (k/2-1)\ln \hat r - \bar \mu_0\Big)^2 + 2^{-t} \bar\sigma_0^2, & \text{ if } \delta=1/2
	\end{cases}$$
	Further, assuming the constraint continues to be inactive and that $\delta \leq \frac{1}{2}$, the system converges exponentially to the steady state
	$\mathcal{L}_t \rightarrow \mathcal{D}\hat r$,
	$\bar \mu_t \rightarrow 0$,
	$\bar\sigma_t^2 \rightarrow 0$.
\end{theorem}

\begin{center} \hyperlink{pf:stable2}{\texttt{[Link to Proof]}} \end{center}

Notice that if the leverage constraint in the system is reached, we can still treat the system as a reset of $\bar\mu_0$ and $\bar\sigma_0$ when we reach a point at which the constraint is no longer binding. While the system subsequently remains without a binding constraint, we again converge to a steady state starting from the new initial conditions.

\paragraph{Interest rates and trading below \$1.}
A consequence of Theorem~\ref{prop:stable2} is that the DStablecoin will trade below target during times in which Ether expectations are high. This is empirically seen in Figure~\ref{fig:dai_below_target}. An interest rate charged to speculators can balance the market (the `stability fee' in Dai). This can temper expectations by effectively reducing $r$ in Theorem~\ref{prop:stable2}. In the stable steady state, setting the interest rate to offset the average expected ETH return will achieve the price target. However, this is practically difficult as $r$ changes over time and is difficult to measure accurately. It also depends on holding periods of speculators. It is an open question how to target these fees in a way that maintains long-term stability.

\subsection{Instability if leverage constraint is binding}\label{sec:instability}
When the speculator's leverage constraint is binding, DStablecoin price behavior can be more extreme. We argue informally that this can lead to high volatility in our model. The probability distribution for the leverage constraint to be binding in the next step has a kink at the boundary of the leverage constraint. In particular, it becomes increasingly likely that the leverage constraint is binding in a subsequent step due to deleveraging effects described previously. Note that feedback of large liquidations on Ether price, if added to the model, will add to this effect.

\begin{figure}
	\centering
	\begin{subfigure}[b]{0.49\textwidth}
		\includegraphics[width=\textwidth]{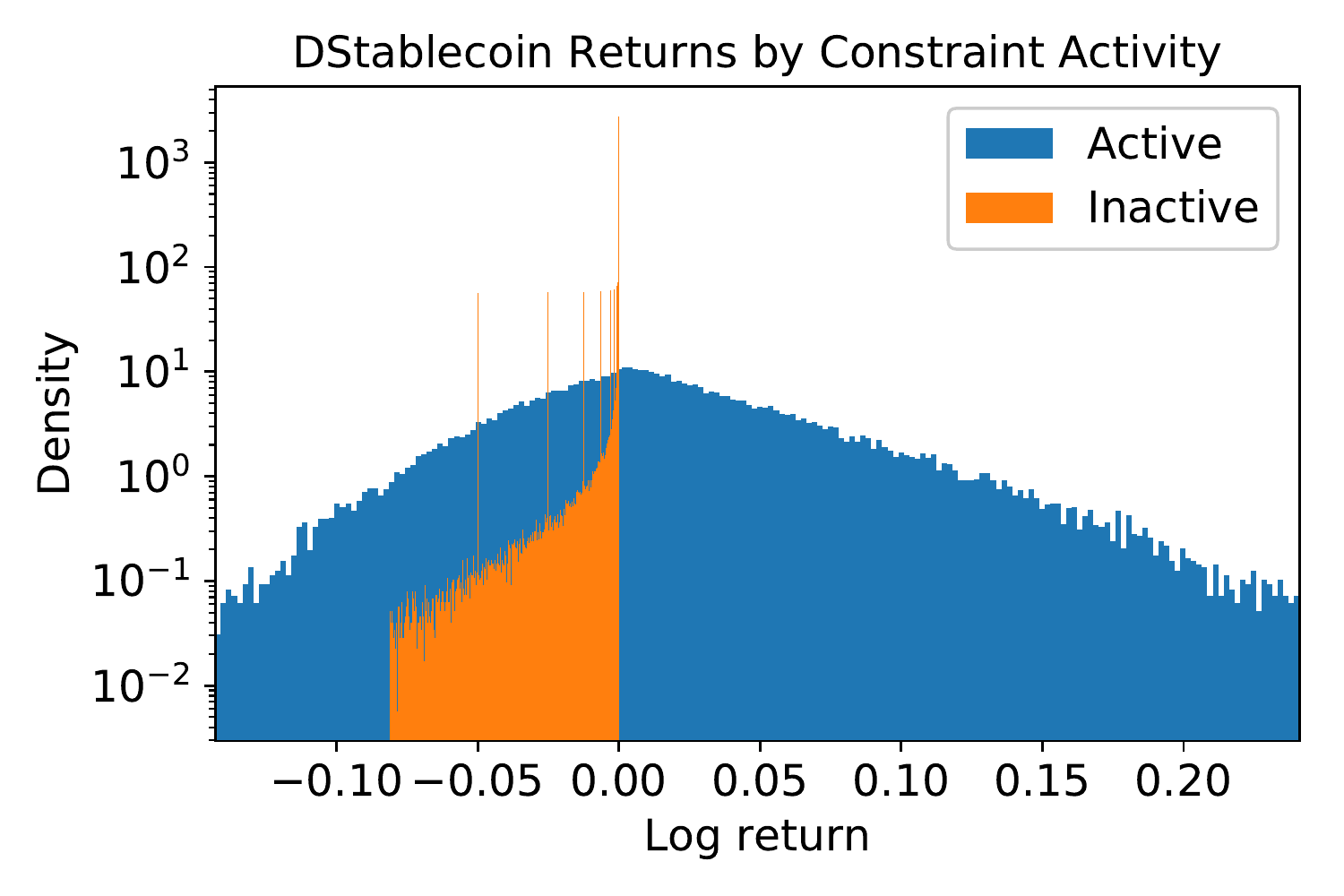}
		\caption{Histogram of DStablecoin returns when leverage constraint is binding vs. non-binding with constant $\hat r$.}\label{fig:hist_constraint_returns}
	\end{subfigure}
	\begin{subfigure}[b]{0.49\textwidth}
		\includegraphics[width=\textwidth]{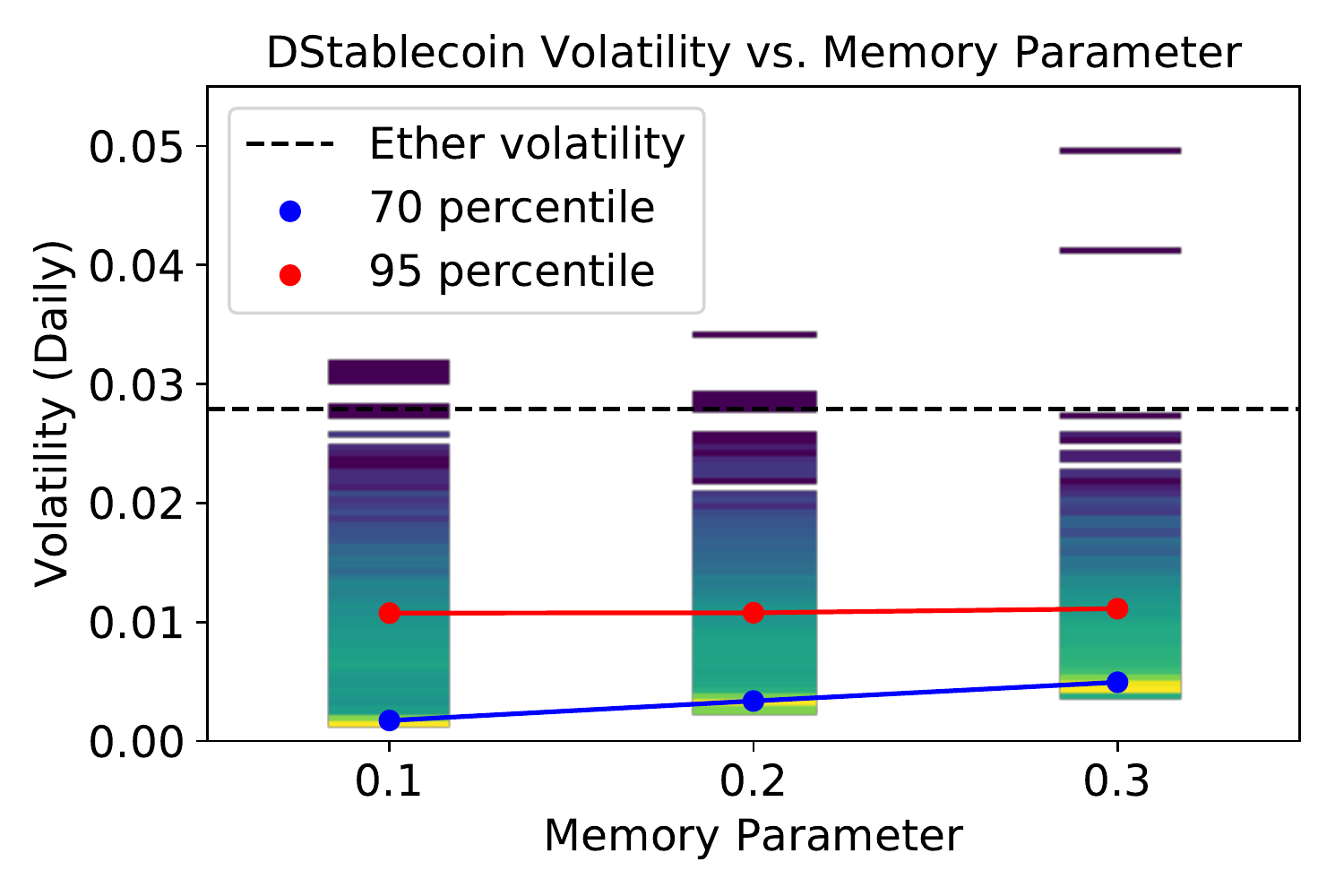}
		\caption{Heat map of volatility under different speculator $\gamma=\delta$ memory parameters.}\label{fig:hist_vol_learning_rate}
	\end{subfigure}
	\caption{DStablecoin volatility, 10k simulation paths of length 1000.}
\end{figure}

We show such instability computationally in Figure~\ref{fig:hist_constraint_returns} in simulation results. In this figure, the shape of the inactive histogram reflects the speculator's willingness to sell at a slight discount when the leverage constraint is non-binding due to the constant $\hat r$ assumption.

We relax this assumption in Figure~\ref{fig:hist_vol_learning_rate}, which shows the effects on volatility of different speculator memory parameters. This figure is a heat map/2D histogram. A histogram over $y$-values is depicted in the third dimension (color: light=high density, dark=low density) for each $x$-value. Each histogram depicts realized volatilities across 10k simulation paths using the simulation setup introduced in the next section and the given memory parameter ($x$-value). Horizontal lines depict selected percentiles in these histograms. The dotted line depicts the historical level of Ether volatility for comparison.

In Figure~\ref{fig:hist_vol_learning_rate}, volatility is bounded away from 0 even in non-binding leverage constraint scenarios; the distance increases with the memory parameter. This happens because $r$ updates faster with a higher memory parameter. As the speculator's objective then changes at each step, the steady state itself changes. Thus we expect some nonzero volatility, although it remains low in most cases.

In not-so-rare cases, however, volatility can be on the order of magnitude of actual Ether volatility in these simulations. As seen in Figure~\ref{fig:vol_risk_mgmt}, this result is robust to a wide range of choices for the speculator's risk constraint. This suggests that DStablecoins perform well in median cases, but are subject to heavy tailed volatility.

\section{Simulation Results}\label{sec:simulations}

We now explore simulation results from the model considering a wide range of choices for the speculator's risk constraint. Unless otherwise noted, the simulations use the following parameter set with a simplified constant demand assumption ($\mathcal{D}=100$) and a t-distribution with df=3 to simulate Ether log returns. This carries over the simplified model from Section~\ref{sec:stable_v_unstable}, although other choices are also amenable to simulation. Cryptocurrency returns are well known for having very heavy tails. This choice gives us these heavy tails with finite variance. Note, however, that this doesn't capture path dependence of Ether returns. We instead assume Ether returns in each period are independent. We run simulations on 10k paths of 1000 steps (days) each. This is enough time to look at short-term failures and dynamics over time. The simulation code is available with full details at \url{https://github.com/aklamun/Stablecoin_Deleveraging}.

\begin{center}
	\begin{tabular}{c|l|l}
		\textbf{Parameter}	&	\textbf{Value}	& \textbf{Rationale} \\
		\hline
		$n_0$		&	$400$ 	&	4x initial collateralization $>$ typical Dai level \\
		$r_0$		&	$1.00583$	& Historical daily Ether mult. return 2017-2018 \\
		$\mu_0$		&	$0.00162$	& Historical daily Ether log return 2017-2018 \\
		$\sigma_0$	&	$0.027925$	& Historical daily Ether volatility 2017-2018 \\
		$\gamma=\delta$	&	$0.1$	& $\sim$ Recommended value \cite{longerstaey1996} \\
		$\beta$		&	$1.5$	& Threshold used in MakerDAO's Dai \\
		$\alpha$	&	$\sim 1.28$	& Value assuming normal distr. + $a=0.1$ \\
		$b$		&	$1$	& Consistent with VaR constraint \\
	\end{tabular}
\end{center}

Note that our simulations study daily movements. We choose this time step to examine these systems under reasonable computational requirements. More realistic simulations might study intraday movements. One plausible scenario of a Dai freeze is if the price feed moves too far too fast instraday, so that speculators don't have enough time to react before liquidations are triggered and keepers (who perform actual liquidations) are unable to handle the avalanche of liquidations. As the price feed in Dai faces an hourly delay in the price feed, hourly time steps are a natural choice for follow-up simulations. This said, daily time steps can actually be reasonable due to a behavioral trend in Dai data: most Dai speculators realistically don't track their positions with very high frequency as supported by overall high liquidation rates.

\subsection{Speculator behavior affects volatility}
We compare DStablecoin performance under the following speculator behaviors encoded in the risk constraint.
\begin{center}
	\begin{tabular}{c|l}
		\textbf{Name}	&	\textbf{Speculator risk constraint} \\
		\hline
		VaRN.1	&	 VaR using $a=0.1$ + normality assumption \\
		VaRN.01	&	VaR using $a=0.01$ + normality assumption \\
		VaRM.1	&	VaR using $a=0.1$ + heavy-tailed assumption \\
		VaRM.01	&	VaR using $a=0.01$ + heavy-tailed assumption \\
		AC1		&	Anti-cyclic constraint, $b=-0.5$, $\alpha=0.01$ \\
		AC2		&	Anti-cyclic constraint, $b=-0.5$, $\alpha=0.02$ \\
		RN		&	Risk neutral, only faces liquidation constraint
	\end{tabular}
\end{center}

\begin{figure}
	\centering
	\begin{subfigure}[b]{0.49\textwidth}
		\includegraphics[width=\textwidth]{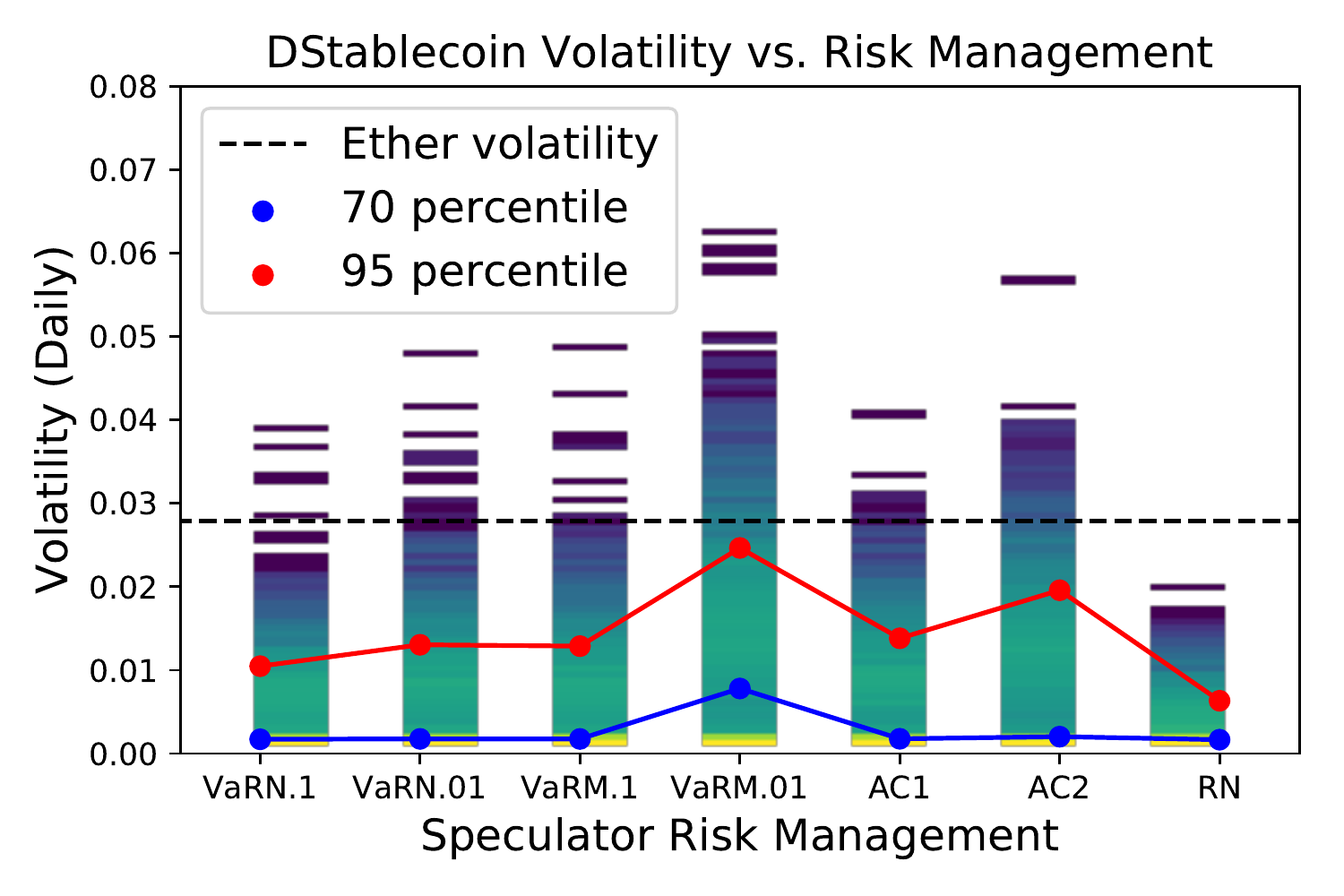}
		\caption{Ether returns$\sim\text{t-distr}(\text{df}=3,\mu=0)$}\label{fig:hist_vol_risk_mgmt_tdist}
	\end{subfigure}
	\begin{subfigure}[b]{0.49\textwidth}
		\includegraphics[width=\textwidth]{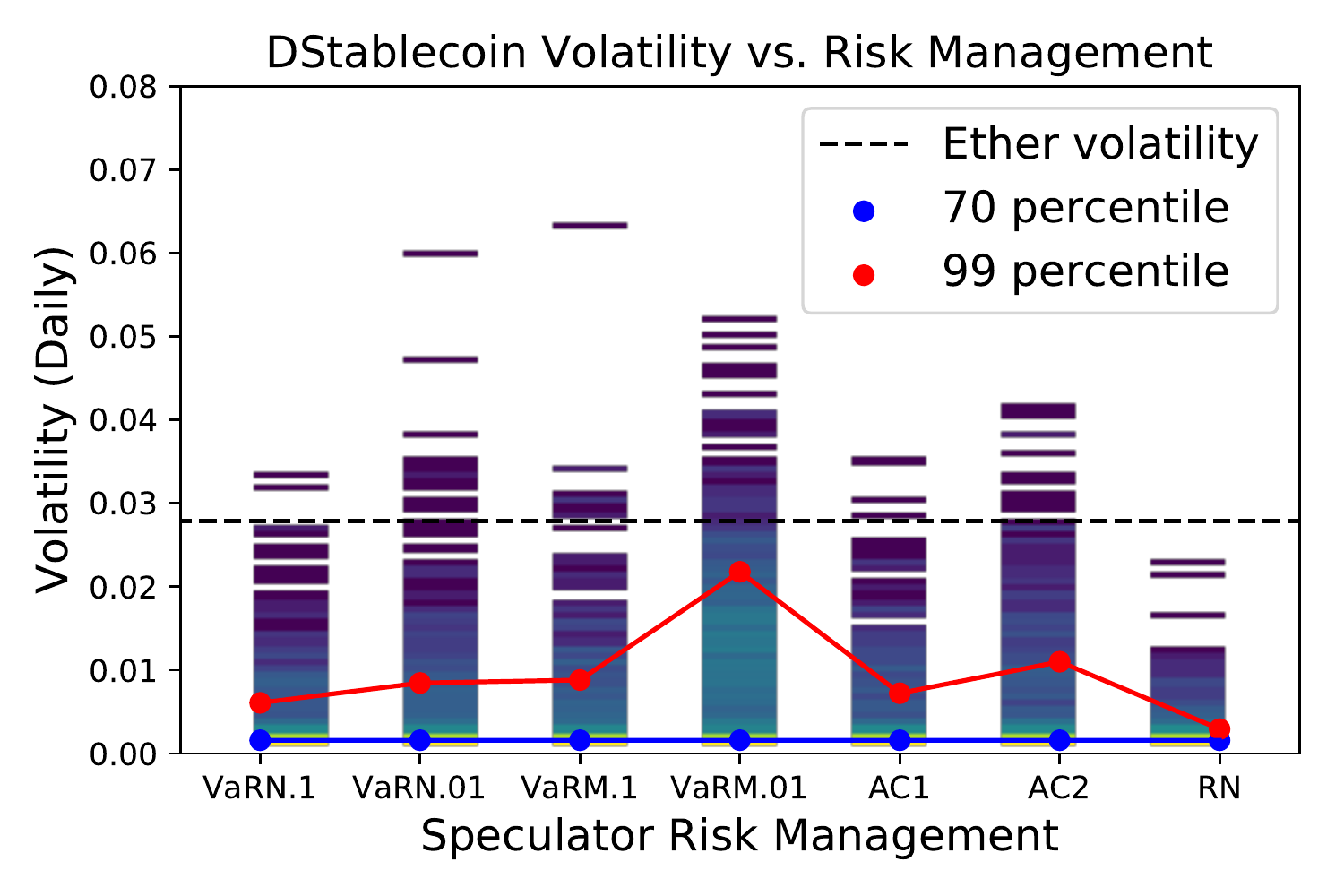}
		\caption{Ether returns$\sim\text{t-distr}(\text{df}=3,\mu=r_0)$}\label{fig:hist_vol_risk_mgmt_drift_nz}
	\end{subfigure}
	\begin{subfigure}[b]{0.49\textwidth}
		\includegraphics[width=\textwidth]{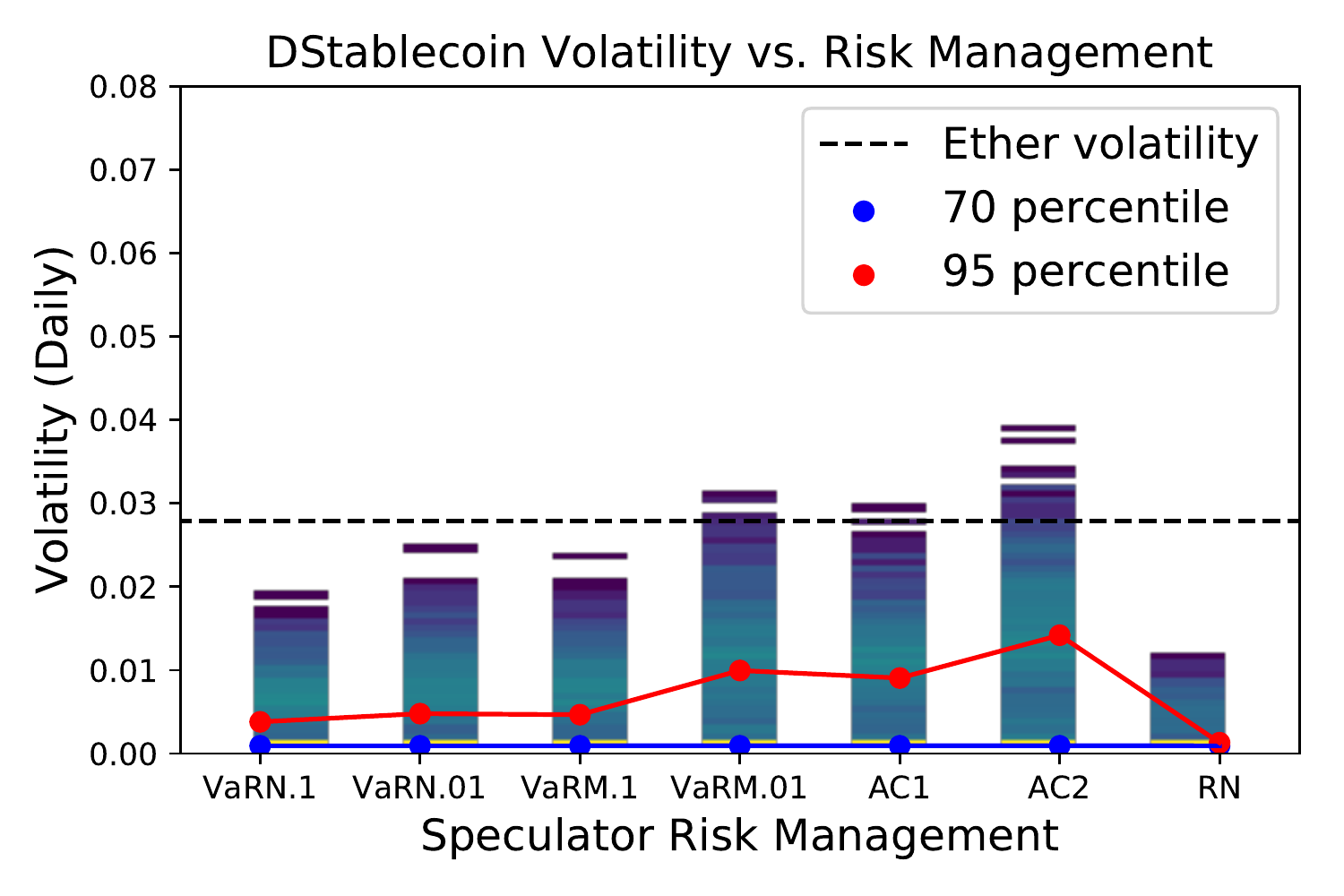}
		\caption{Ether returns$\sim\text{normal}(\mu=0)$}\label{fig:hist_vol_risk_mgmt_normal}
	\end{subfigure}
	\caption{Heatmaps of DStablecoin volatility for different speculator risk management behaviors.}\label{fig:vol_risk_mgmt}
\end{figure}

Figure~\ref{fig:vol_risk_mgmt} compares the effects on volatility of these behavioral constraints under various Ether return distributions. These figures are heatmaps/2D histograms similar to that in Figure~\ref{fig:hist_vol_learning_rate}. The results suggest that DStablecoins face significant tail volatility (on the order of Ether volatility) even under comparatively `nice' assumptions on Ether return distributions, such as with significant upward drift (Figure~\ref{fig:hist_vol_risk_mgmt_drift_nz}) and a normal distribution (Figure~\ref{fig:hist_vol_risk_mgmt_normal}). Figure~\ref{fig:simulation_msd} depicts relative (\% difference) mean-squared difference of simulated volatility for the different risk management methods vs. a risk neutral speculator. The mean-squared difference is large, suggesting that the speculator's risk management method has a large effect on volatility.

The results suggest how speculator behavior can affect DStablecoin volatility within the model. Stricter cyclic risk management (e.g., VaR) on the part of the (single) speculator can lead to increased DStablecoin volatility without improving the safety of the system. Whether countercyclic (setting constraint to increase leverage during downturns) or cyclic (setting constraint to decrease leverage during downturns), the resulting DStablecoin volatility is connected with how narrow the feasible region for the constraint becomes. A risk neutral speculator, which has the widest feasible region for the constraint, leads to the lowest volatility. Stricter risk management serves to reduce the feasible region. Note that these results may be different if there are multiple types of speculators, for instance some that are cyclic and others that are countercyclic.

Figure~\ref{fig:hist_vol_learning_rate} further suggests that a higher speculator memory parameter (lower memory) tends to increase volatility in typical cases. This makes sense as high memory parameters can lead to noise chasing on the part of the speculator. Note that keeping the speculator's expected Ether returns and variance constant is equivalent to setting a static risk constraint.

\subsection{Stable asset failure is dominated by collateral asset returns}
We define the DStablecoin's \textbf{failure (or stopping) time} to be either (1) when the speculator's liquidation constraint is unachievable or (2) when the DStablecoin price remains below \$0.5 USD. In these cases, a global settlement would be reasonable, leaving DStablecoin holders with Ether holdings with high volatility in subsequent periods.

\begin{figure}
	\centering
	\begin{subfigure}[b]{0.49\textwidth}
		\includegraphics[width=\textwidth]{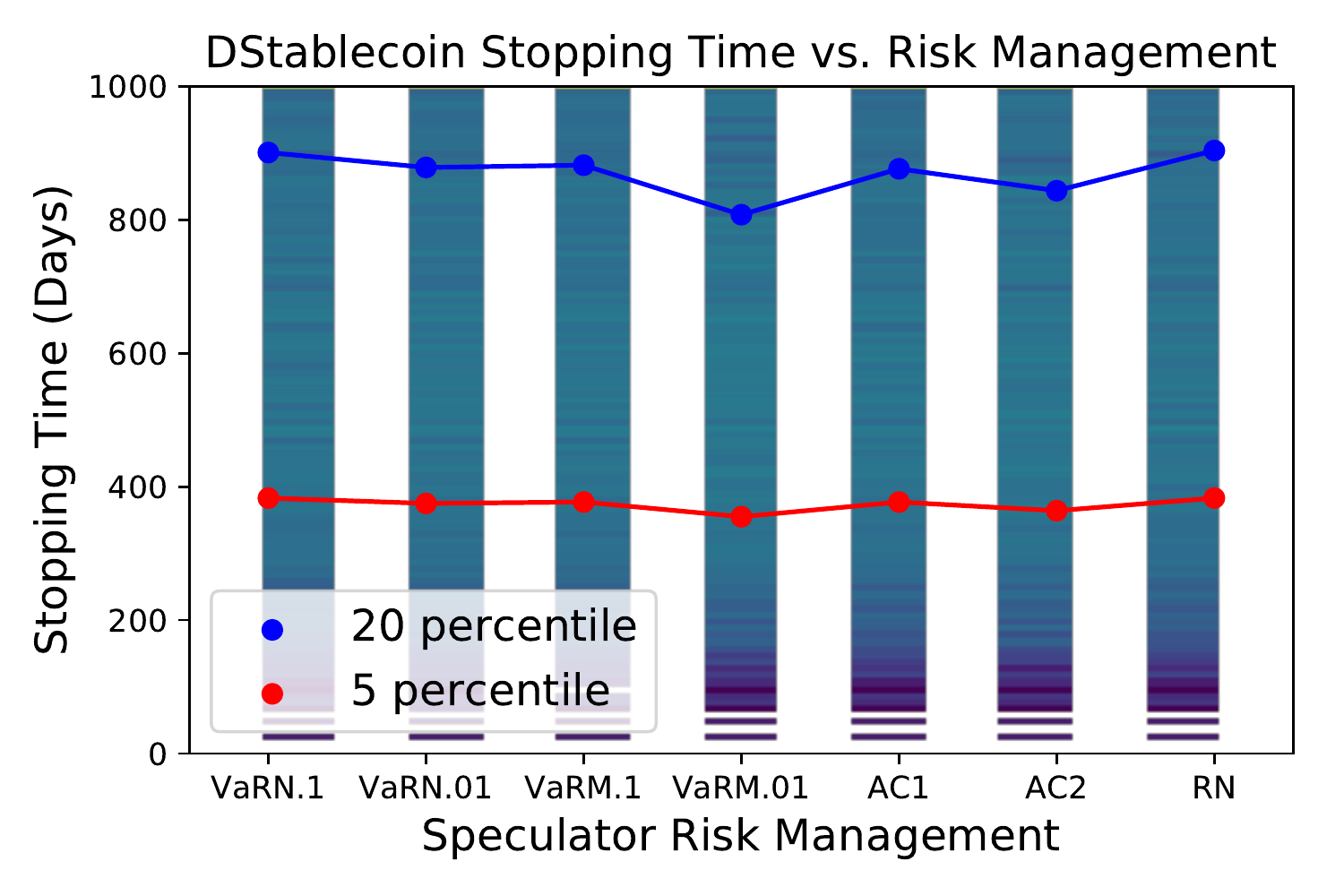}
		\caption{Ether returns$\sim\text{t-distr}(\text{df}=3,\mu=0)$}\label{fig:hist_stopping_risk_mgmt_tdist}
	\end{subfigure}
	\begin{subfigure}[b]{0.49\textwidth}
		\includegraphics[width=\textwidth]{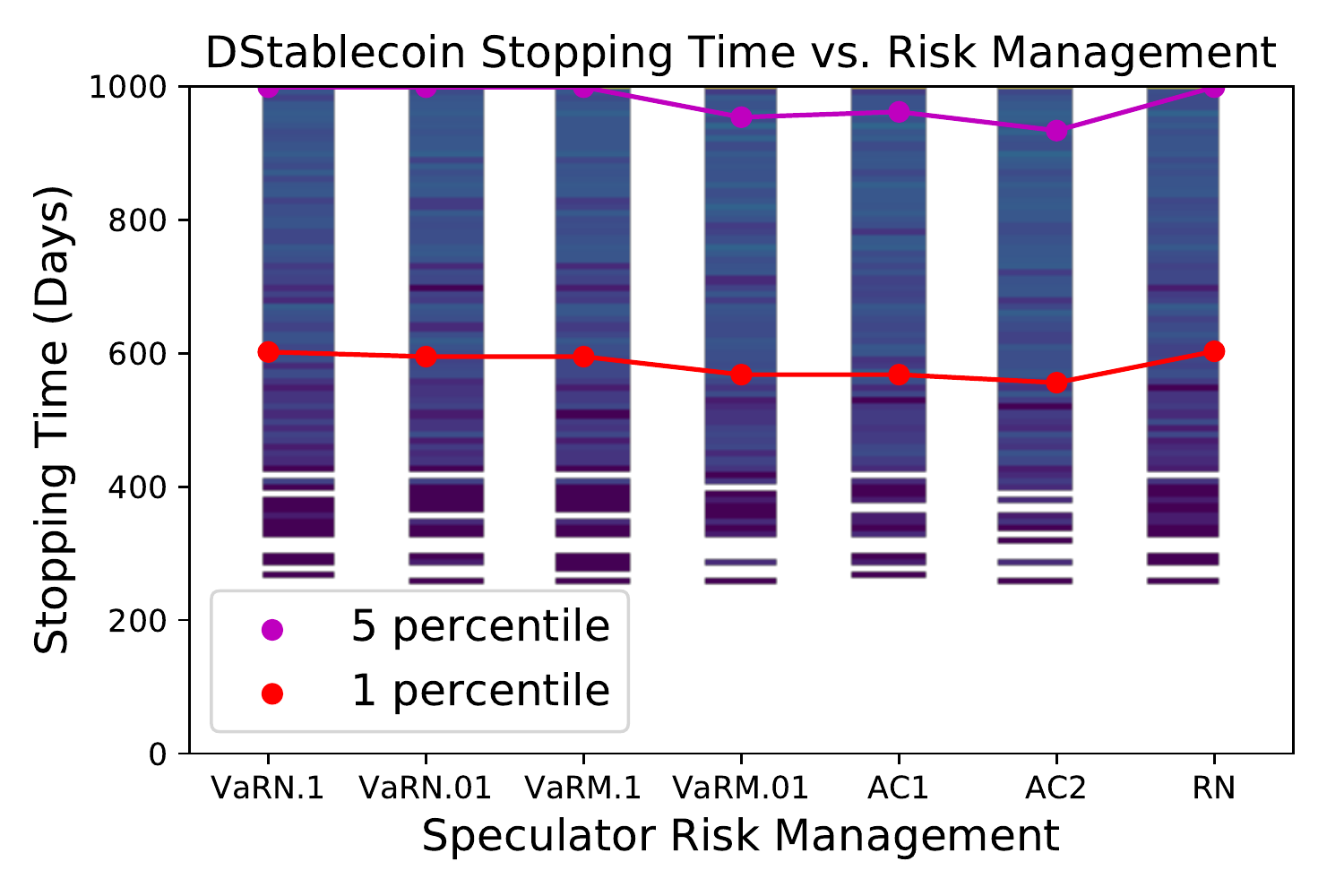}
		\caption{Ether returns$\sim\text{normal}(\mu=0)$}\label{fig:hist_stopping_risk_mgmt_normal}
	\end{subfigure}
	\caption{Heatmaps of DStablecoin failure times for different speculator risk management behaviors.}\label{fig:stopping_risk_mgmt}
\end{figure}

\begin{figure}
	\centering
	\includegraphics[width=0.7\textwidth]{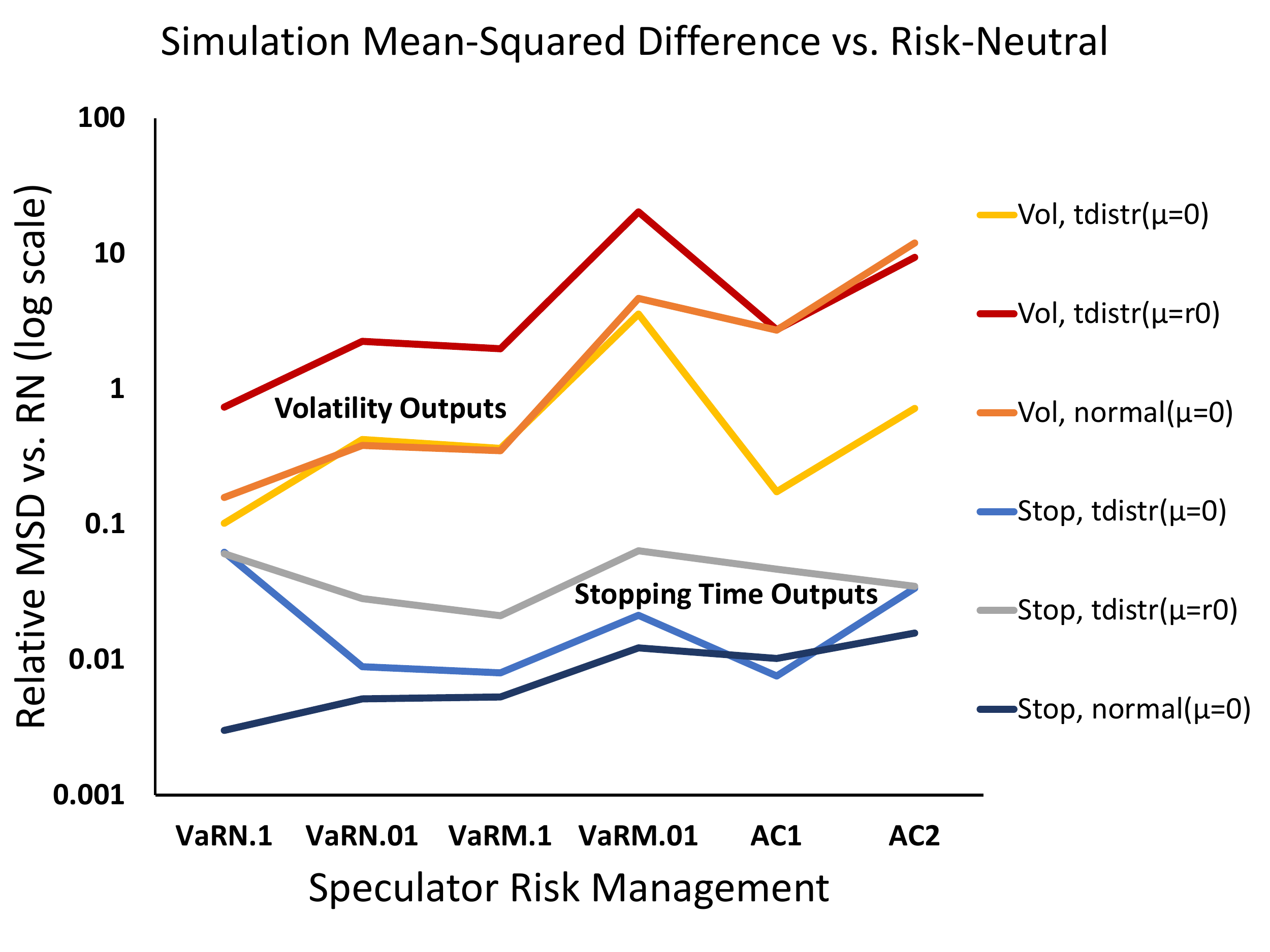}
	\caption{Relative mean-squared difference (MSD) of simulated volatility and stopping time for given speculator strategy vs. risk neutral strategy. Different lines represent different output (volatility or stopping time) and different return distribution assumptions for the simulations.}\label{fig:simulation_msd}
\end{figure}

Figure~\ref{fig:stopping_risk_mgmt} compares the effects on failure time of these behavioral risk constraints. The stopping time distributions appear comparable across a wide range of selections for the speculator's risk constraint. They are additionally comparable across the memory parameters studied above. Figure~\ref{fig:simulation_msd} depicts relative mean-squared difference of simulated stopping times for the different risk management methods vs. a risk neutral speculator. In calculating the mean-squared difference, we only include cases in which the failure is realized within the simulation. The mean-squared difference is small (1-2 orders of magnitudes smaller than for volatility), providing additional evidence that the stopping time is largely independent of the speculator's risk management. In particular, a large proportion of failure events would not have been prevented by different speculator risk management within the model.

DStablecoin failure probabilities appear to be dominated by Ether returns as opposed to speculator behavior. The results suggest that DStablecoins may not be long-term stable, even under comparatively `nice' assumptions for Ether return distributions. To avoid failure, they would essentially rely on more speculator capital entering the system during downturns.

\section{Stablecoin Attacks}\label{sec:attacks}

Attacking a DStablecoin is different than traditional currency attacks. The focus is not on breaking the willingness of the central bank to maintain a peg. It instead involves manipulating the interaction of agents. We show that stablecoin design can enable profitable trades against stability that attack the system. These come from the existence of profitable trades around liquidations and the ability of miners to reorder and censor transactions to extract value.

\subsection{Expanded Model: Adding an Attacker}

We consider an expanded model under the fixed outside demand setting of the previous section. In the expansion, we consider an attacker, who can speculatively enter/exit the DStablecoin market. The attacker can buy $\delta$ dollar-value of DStablecoin at some time $t$ with the goal of selling it at a later time $s$ for $\delta + \varepsilon$. These occurrences change the demand structure: $\mathcal{D}_t = \mathcal{D} + \delta$, $\mathcal{D}_s = \mathcal{D} - (\delta + \varepsilon)$.

\subsection{Profitable bets on liquidations}

Table~\ref{table:attack} illustrates an example scenario for a profitable bet on liquidations. The attacker injects $\delta = 1$ in demand at $t=1$, which acquires $1.0008$ DStablecoins at $p_1^D$. In $t=3$, after the liquidation, the attacker is then able to extract $\delta +\varepsilon = 1.083$ from selling the DStablecoin. This yields a return of $8.3\%$. This is akin to a short squeeze on existing speculators. It takes advantage of the fact that liquidations occur at DStablecoin market rate, which in turn affects the market rate.
\begin{table}
	\centering
	\begin{tabular}{c|c|c|c|c|c|c|c}
		$t$ & $p_t^E$ & $\delta + \varepsilon$ & $\mathcal{D}_t$ & $\Delta_t$ & $\mathcal{L}_t$ & $p_t^D$ & $n_t$ \\
		\hline
		$0$ & $85$ & & $100$ & & $100.583$ & $0.994$ & $1.8$ \\
		$1$ & $85$ & $+1$ & $101$ & $0.502$ & $101.085$ & $0.999$ & $1.806$ \\
		$2$ & $82$ & & $101$ & $-8.716$ & $92.369$ & $1.093$ & $1.690$ \\
		$3$ & $82$ & $-1.083$ & $99.917$ & & $92.369$ & $1.082$ & $1.689$ \\
	\end{tabular}
	\caption{Example scenario of a profitable bet on liquidations.}\label{table:attack}
\end{table}

The attacker can do better by choosing $\delta,\varepsilon$ to maximize $\varepsilon$ subject to $\frac{\delta + \epsilon}{p_2^D} \leq \frac{\delta}{p_o^D}$. Choosing $\delta=4.5, \varepsilon=0.59$ (not optimal) yields a return of $13\%$. The attacker could also spread out $\delta$ over a longer period of time to achieve lower purchase prices.

From a practical perspective, the optimization is sensitive to misestimation of demand elasticity. While Dai has hit prices as high as \$1.37 historically (source: coinmarketcap), it hasn't typically reached prices above \$1.09. Thus smaller bets (relative to supply) may be safer. Regardless, these can be large opportunities in large systems. In addition, outside of this model, real implementations create arbitrage of $5-13\%$ to automate liquidations.

\subsection{Attacks}

\paragraph{Attack 1:} An attacker bets on an ETH decline and manipulates the market to trigger and profit from spiraling liquidations. This uses the short squeeze-like trades in the previous example. It can also be supplemented with a bribe to miners to freeze collateral top-ups. The attacker could also enter as a new speculator at the high DStablecoin prices after the attack and thus leverage up at a discount. Outside of the model, the attack may have a negative effect on the long-term DStablecoin demand due to the induced volatility. This can be further beneficial to the attacker, who can then also deleverage in the future at a discount.

\paragraph{Attack 2:} The attacker is also a miner and reorganizes the recent transaction history (such as by initiating a fork) to be on the receiving end of arbitrage oppotunities from liquidations. For instance, following an ETH decline, the miner could trigger and profit from spiraling liquidations. In a fork, the attacker creates a new timeline that inherits the ETH price trajectory (via oracle transactions). The attacker can then censor speculator transactions (e.g., collateral top-ups) to trigger new liquidations and extract profit around all liquidations, which are guaranteed in the timeline. If the stablecoin system is large, the miner extractable value can be large (and is additive with other sources of extractable value). This creates the perverse incentive for miners to perform this attack if the attack rewards are greater than lost mining rewards. This is similar to the time-bandit attack in \cite{daian2019}.

\vspace{0.3cm}
In Attack 1, the attacker takes on market risk as the payoff relies on a future ETH decline and liquidation. It is a speculative attack that can induce volatility in the stablecoin. In Attack 2, the attacker's payoffs are guaranteed if the attack fork is successful. These payoffs incentivize blockchain consensus attack. A possible equilibrium is for miners to collude and share this value.

These attacks occur in a permissionless setting, in which agents can enter/exit at any time with a degree of anonymity. While in traditional finance, market manipulation rules can be enforced legally, in decentralized finance, enforcement is only possible to the extent that it can be codified within the protocol and incentive structure. We leave to future study a full exploration of these incentive structures in a game theoretic setting based on foundations for blockchain forking models set in, e.g., \cite{biais2018}.

Since the initial release of this paper, this attack surface around stablecoin liquidations was exploited in related ways to Attack 2. In Attack 2, a miner reorganizes the recent history to extract profit from arbitrage opportunities from liquidations. In reality on Black Thursday, mempool manipulation contributed to the clearing of \$8m of Dai liquidation auctions at near zero prices \cite{blocknative2020}.

\paragraph{Mitigations.} We discuss some preliminary ideas toward mitigating attack potential. Liquidations could be spread over a longer time period. This could potentially lessen deleveraging spirals by smoothing demand and increase the costs to a forking attack. However, it presents a trade-off in that slow liquidations come with higher risks to the stablecoin becoming under-collateralized. We also suggest tying oracle prices and DEX transactions to recent block history so that a reorganization attack can't easily inherit price and exchange history. Practically, however, this may be difficult to tune in a way that's not disruptive as small forks happen normally.

\section{Discussion}\label{sec:discussion}

In general, it is impossible to build a stablecoin without significant risks. As speculators participate by making leveraged bets, there is always an undiversifiable cryptocurrency risk. However, a stablecoin can aim to be an effective store of value assuming the cryptocurrency market as a whole is not undermined. In this case, it is \emph{conceivable} to sustain a dollar peg if the stablecoin survives transitory extreme events. That is, to achieve long-term probabilistic stability, a stablecoin should maintain a high probability of survival.

\paragraph{Failure risks.} DStablecoins are complex systems with substantial failure risks. Our model demonstrates that they can work well in mild settings, but may have high volatility outside of these settings. As we explore in this paper, the market can collapse due to feedback effects on liquidity and volatility from deleveraging effects during crises. These effects can exacerbate collateral drawdown. Surviving these events may rely on bringing in increasing amounts of new capital to expand the DStablecoin supply during such crises. In these events speculators may not always be willing and able to take these new risky positions. Indeed, there are may examples of speculative markets drying up during extreme market movements. As we explore below, continued stability during these events additionally relies on new capital entering the system \emph{in a well-behaved manner} as profitable attacks are possible.

As suggested by our simulations, stablecoin holders face the direct tail risk of cryptocurrencies. If the market loses liquidity, there is no guarantee that forced liquidation of speculators' collateral will be possible within reasonable pricing limits. Further, volatile cryptocurrency markets can, in unlikely events, move too fast for speculators to adapt their positions. In these cases, stablecoin holders can only truly rely on the cryptocurrency value from global settlement.

\paragraph{Remark on oracle risks.}
The DStablecoin design also relies on trusted oracles to provide real world price data, which could be subject to manipulation. In MakerDAO's Dai, for instance, oracles are chosen by MKR token holders, who vote on system parameters. This opens a potential 51\% attack, in which enough speculators buy up MKR tokens, change the system to use oracles that they manipulate, and trigger global settlement at unfavorable rates to stablecoin holders while pocketing the difference themselves when they recover their excess collateral. A hint of manipulation in oracles or large acquisitions of MKR could potentially trigger market instability issues on its own.

Note that Dai has protections from oracle attacks.\footnote{Though it is notable that most MKR is reputedly held by just a few individuals within the MakerDAO team.} First, there is a threshold of maximum price change and an hourly delay on new prices taking effect. This means that emergency oracles have time to react to an attack. Second, at current prices 51\% of MKR is substantially more expensive than the ETH collateral supply. However, this second point does not have to be true in general--at least unless Dai holders otherwise bid up the price of MKR for their own security. The value of MKR is linked to expectations around Dai growth as fees paid in the system are used to reduce MKR supply. At some point, the expectation may not be enough to lift MKR value above collateral on its own. This raises the question of whether fees should be used to reduce MKR supply at all. Alternatively, MKR value could be completely based on the potential value of a 51\% attack, which may also grow with Dai growth, and the value of fees could be put to different uses, as we discuss further below.

\paragraph{A good fee mechanism may quell deleveraging spirals.}
Dai imposes fees on speculators when they liquidate positions (e.g., liquidation penalty, stability fee, penalty ratio). These can \emph{amplify} deleveraging effects by increasing deleveraging costs and disincentivizing new capital from entering the system during crises. An alternative design with automatic counter-cyclic fees could enhance stability by reducing feedback effects. For instance, fees could be collected while the system is performing well, but these fees could be removed (or made negative) automatically during liquidity crises in order to limit feedback effects and remove disincentives to bringing new capital into the system.

Speculators in Dai can pay back liabilities at any time and come and go from the system, which raises concerns about herd behavior in crises. A herd trying to deleverage can trigger a deleveraging spiral. Dynamic fees tuned to inflow/outflow could additionally disincentivize herd behavior to deleverage at the same time.

\paragraph{An alternative `collateral of last resort' idea in Dai.}
In Dai, MKR serves a certain `last resort' role in addition to governance. If there is a collateral shortfall, then new MKR is minted and sold to cover Dai liabilities making up the shortfall. This may not always be possible as the MKR market can similarly face illiquidity and the market cap may not be high enough to cover shortfalls. In some settings, MKR holders might actually have an incentive to trigger a global settlement early before MKR would be inflated. A Dai shutdown would have some effect on the price of MKR, but the cost may be small if MKR holders expect a successful relaunch of Dai after the crisis. An early shutdown is not ideal for Dai holders, as they will want to hold the stable asset for longer during extreme events. In addition to incentive alignment being unclear in MKR's `last resort' role, the invocation of the role only helps cover the aftermath of a crisis (an existing shortfall) as opposed to quelling the effects that cause the crises.

We propose an alternative `last resort' role of governance tokens that instead aims to quell deleveraging spirals. This could be achieved by automatically positioning the MKR supply as system collateral against which Dai can be minted to expand supply in crises. To illustrate, if there is a massive deleveraging by speculators, leading to excess demand for Dai and an inflated Dai price, then new Dai could be automatically minted against the MKR supply as collateral to help balance the market. In this way, a deleveraging spiral is damped: should a new wave of speculator deleveraging be triggered, it will not compound the price effect from the past wave. System fee revenue could also be put to this use.

\paragraph{Uses of limited fee revenue.}
Dai produces limited fee revenue, most of which rewards MKR investors. There is additionally a Dai savings rate that rewards Dai holders using fee revenue and serves as another tool to balance the Dai market (e.g., to boost demand for Dai when the price is below target). There is an inherent trade-off in using fee revenue, however. A Dai savings rate uses this revenue to improve stability in relatively normal settings in which a higher fee itself serves to balance the market. Alternatively, fee revenue can be channeled to an emergency fund that lessens the severity of crises--for instance as suggested above. These fees and their potential uses can be incorporated into our model to compare the effects of different design choices.

\paragraph{Stablecoin risk tools.}
Our results suggest tools and indicators that can warn about volatility in DStablecoins. We can find proxies for the free supply, estimate the price impact of liquidations, and track the entrance of new capital into speculative positions. We can connect this information with model results to estimate the probability of liquidity problems given the current state. This information is also useful in valuing token positions in these systems (e.g., Dai, MKR, and the speculator's leveraged position). 

Some exchanges have bundled select stablecoins into a single market that ensures 1-to-1 trading (e.g., \cite{huobi2018}). In this case, exchanges are essentially providing insurance to their users against stablecoin failures. These arrangements could lead to a run on exchanges in the event that some stablecoins fail. It is unclear if these exchanges are subject to regulation to protect users against this, and it is further unclear if such regulations would be sufficient to account for risks in stablecoins. Our model provides insight into the risks (to exchanges and users) if such arrangements in the future include non-custodial stablecoins.

\paragraph{Future directions.}
We suggest expansions to our model to explore wider settings.
\begin{itemize}
	\item Incorporate more speculator decisions, such as locking and unlocking collateral and holding different assets, accommodating speculators with security lending motivation. This makes the speculator's optimization problem multi-dimensional. In this expanded setting, speculators may make more long-term strategic decisions considering whether tomorrow they would have to buy back stablecoins and at what price.
	\item Consider multiple speculators with different utility functions who participate in the DStablecoin market. In this expanded setting, we can consider the conditions under which new capital may enter the system and formally study the economic attack described above and the effects of external incentives.
	\item Incorporate additional assets, such as a custodial stablecoin that faces counterparty risk. This would allow us to study long-term movements between stablecoins in the space and learn about systemic effects that could be triggered by counterparty failures. This is further relevant in evaluating systems like Maker's multi-collateral Dai. However, this comes with a trade-off of a new counterparty risk that is very hard to measure. In particular, it's not just custodian default risk, but also risk of targeted interventions on centralized assets. Such interventions (e.g., from a government who wants to shut down Dai) could be highly correlated with cryptocurrency downturns as that is when the system is naturally weakest.
	\item Incorporate endogenous feedback of liquidations on Ether price, which becomes relevant if the DStablecoin system becomes large relative to the Ether market. This is similarly important for \emph{endogenous collateral} stablecoins like Synthetix sUSD and Terra UST, in which a system equity-like asset is used as collateral (see \cite{klagesmundt2020stablecoins}).
\end{itemize}
Additionally, our existing model can be adapted to analyze DStablecoins with different design characteristics. For instance,
\begin{itemize}
	\item DStablecoins with more general collateral settlement, in which stablecoin holders can individually redeem stablecoins for collateral. This is possible, for instance, in bitUSD and Steem Dollars, and more recently in Celo Dollars. In this case, the stablecoin acts as a perpetual option to redeem collateral, and stablecoin volatility will be additionally related to the settlement terms.
	\item DStablecoins without speculator agents (e.g., Steem Dollars, in which the whole marketcap of Steem acts as collateral, or Celo Dollars, in which Celo reserves act as collateral). In these systems, stablecoin issuance is automated with the rest of the protocol. Our model can be adapted by removing speculator decisions and modeling the growth of collateral from block rewards and growth of stablecoin from other processes.
	\item Some non-collateralized algorithmic stablecoins. We believe this setting can also be interpreted in our model by thinking of \emph{implicit collateral} that ends up describing user faith in the system (see \cite{klagesmundt2020stablecoins}). The underlying mechanics would be similar, simply recreating `out of thin air' the value of the underlying asset as opposed to building on top of the value of an existing asset. The stability of the system ultimately still relies on how people perceive this value over time similarly to how perceived value of Ether changes.
\end{itemize}

\begin{acks}
We thank David Easley, Steffen Schuldenzucker, Christopher Chen, Akaki Mamageishvili, Peter Zimmerman, Sergey Ivliev, Tomasz Stanczak, Sid Shekhar, as well as the participants of the ECB P2P Financial Systems (2019) workshop, Crypto Valley Conference (2019), and Crytpo Economics Security Conference (2019) for their valuable feedback. This paper is based on work supported by NSF CAREER award \#1653354. AK thanks Lykke, Binance, and Amherst College for additional financial support.
\end{acks}

\bibliographystyle{ACM-Reference-Format}

\appendix
\section{Derivation of Results}

\noindent\rule{\textwidth}{1pt}
\paragraph{Prop.~\ref{prop:constraint_sol}} \hypertarget{pf:constraint_sol}{}
\begin{proof}
	In each period $t$, we determine the leverage constraint by setting $\tilde \lambda = \lambda$ and solving for $\Delta$. Using the formulation of $p^D_t$ from the market clearing, we have the following equation for $\Delta$:
	$$\tilde \lambda \Big(z + \Delta \frac{x}{\Delta - y}\Big) = \beta(\mathcal{L} + \Delta).$$
	Given $\Delta>y$, this transforms to the quadratic equation for $\Delta$
	$$-\beta \Delta^2 + \Delta\Big( \tilde \lambda (z+x) - \beta(\mathcal{L} - y)\Big) - \tilde\lambda zy + \beta\mathcal{L} y =0.$$
	This is a downward facing parabola. The speculator's leverage constraint is satisfied when the polynomial is positive. The roots, if real, bound the feasible region of the speculator's constraint. Due to the requirement that $\Delta > y$, the feasible set is given by $[\Delta_{\min}, \Delta_{\max}] \cap (y, \infty)$. When there are no real roots, the polynomial is never positive, and so the constraint is unachievable.
\end{proof}

\noindent\rule{\textwidth}{1pt}
\paragraph{Prop.~\ref{prop:leverage_sol}} \hypertarget{pf:leverage_sol}{}
\begin{proof}
	By Prop.~\ref{prop:constraint_sol}, $[\Delta_{\min}, \Delta_{\max}] \cap (y, \infty)$ is indeed the feasible region.
	Incorporating the market clearing, the speculator decides $\Delta$ in each period $t$ by solving
	$$\begin{aligned}
	\max \hspace{0.5cm} & r\Big(z + \Delta \frac{x}{\Delta - y}\Big) - \mathcal{L} - \Delta \\
	\text{s.t.} \hspace{0.5cm} & \Delta \in [\Delta_{\min}, \Delta_{\max}] \cap (y, \infty)
	\end{aligned}$$
	
	This optimization is solvable in closed form by maximizing over critical points. Maximizing the objective is equivalent to maximizing
	$$f(\Delta) = r\Delta \frac{x}{\Delta - y} - \Delta.$$
	
	We first consider the case of $\Delta$ approaching $y$ from above and show that this boundary is not relevant in the maximization. The limit is
	$$\lim_{\Delta \rightarrow y^+} f(\Delta) = -\infty.$$
	To see this, note that $\mathcal{L}_{t-1} ~=~ \bar m_{t-1} ~\geq~ w^D_t \bar m_{t-1}$, and so in order to have $\mathcal{L}_t = w^D_t \bar m_{t-1}$, we must have $\Delta<0$. Thus the sign of the term that tends to infinity is negative. The limit is $-\infty$ because the price for the speculator to buy back DStablecoins goes to $\infty$.
	
	To find the critical points of $f$, we set the derivative equal to zero:
	$$\frac{df}{d\Delta} = -\frac{\Delta^2 - 2\Delta y + y(rx +y)}{(\Delta-y)^2}=0$$
	Assuming $\Delta \neq y$, the solutions are the roots to the quadratic
	$\Delta^2 + -2y\Delta + y(rx+y)=0$.
	Notice that the axis of this parabola is at $\Delta=y$. When there are two real solutions, then exactly one of them will be $>y$. Given $y\leq 0$ and $x\geq 0$ and noting $r\geq 0$, a real solution always exists and the relevant critical point is
	$$\Delta^* = y + \sqrt{-yrx}.$$
	
	If it is feasible, $\Delta^*$ is the solution to the speculator's optimization problem. If $\Delta^*$ is not feasible, then we need to choose along the boundary. The possible cases are as follows.
	
	Suppose $\Delta^* < \Delta_{\min}$. Then $\Delta_{\min}$ is feasible since $\Delta^*>y$ implies $\Delta_{\min}>y$. Since $f$ is monotone decreasing to the right of $\Delta^*$, $f(\Delta_{\min})>f(\Delta_{\max})$, and so $\Delta_{\min}$ is the solution.
	
	Suppose $\Delta^* > \Delta_{\max}$. By our assumption that the constraint is feasible, we have that $\Delta_{\max}$ is feasible. Since $f$ is monotone decreasing to the left of $\Delta^*$ on the feasible region, $f(\Delta_{\max})>f(\Delta_{\min})$, and so $\Delta_{\max}$ is the solution.
\end{proof}

\noindent\rule{\textwidth}{1pt}
\paragraph{Prop.~\ref{prop:feasible_condition}} \hypertarget{pf:feasible_condition}{}
\begin{proof}
	The speculator's leverage constraint is unachievable when the quadratic has no real solutions or when all real solutions are $<y$. The first case occurs when
	$$\Big(\tilde \lambda (z+x) - \beta(\mathcal{L}-y)\Big)^2 + 4\beta(-\tilde \lambda zy + \beta \mathcal{L} y) < 0.$$
	
	Noting that
	$y = -w^D \mathcal{L}$ and
	$\mathcal{L} - y = \mathcal{L}(2-w^D)$
	and expanding and simplifying terms yields
	$$\beta \tilde\lambda \mathcal{L} \Big( 2zw^D + 2x(2-w^D)\Big) - (\beta\mathcal{L}w^D)^2 > \Big(\tilde \lambda(x+z)\Big)^2$$
	Completing the square by subtracting $4\beta\tilde\lambda\mathcal{L} x(1-w^D)$ from each side then gives the result.
\end{proof}

\noindent\rule{\textwidth}{1pt}
\paragraph{Prop.~\ref{prop:liquidity_limit}} \hypertarget{pf:liquidity_limit}{}
\begin{proof}
	Setting $z=-\Delta p_t^D = -\Delta \frac{x}{\Delta-y}$ gives the lower bound $\Delta^- := \frac{z}{z+x}y>y$.
	
	Note that $\bar m_t = \mathcal{L}_t$, and so
	$y = \mathcal{L}(w^D - 1) = -w^E \mathcal{L} \leq 0.$
	The term $w^D_t \bar m_{t-1}$ presents a lower bound on the size of the DStablecoin market in the next step from the demand side, and so the speculator can't decrease the size of the market faster than $y$, even with additional capital beyond $z$. As shown above, $\Delta \rightarrow y^+$ coincides with $p^D_t \rightarrow \infty$. The speculator pays increasingly large amounts to buy back more DStablecoins as liquidity dries in the market. 
\end{proof}

\noindent\rule{\textwidth}{1pt}
\paragraph{Prop.~\ref{prop:stable1}} \hypertarget{pf:stable1}{}
\begin{proof}
	With inactive constraint, $\mathcal{L}_t = \sqrt{\mathcal{L}\mathcal{D}\hat r}$,
	$p^D_t = \frac{\mathcal{D}}{\sqrt{\mathcal{L}\mathcal{D}\hat r}} = \sqrt{\frac{\mathcal{D}}{\mathcal{L}\hat r}}$, and
	$\frac{p^D_t}{p^D_{t-1}} = \frac{\sqrt{\frac{\mathcal{D}}{\mathcal{L}\hat r}}}{\frac{\mathcal{D}}{\mathcal{L}}} = \sqrt{\frac{\mathcal{L}}{\mathcal{D}\hat r}}.$
\end{proof}

\noindent\rule{\textwidth}{1pt}
\paragraph{Theorem~\ref{prop:stable2}} \hypertarget{pf:stable2}{}
\begin{proof}
	It is straightforward to verify $\mathcal{L}_t = \mathcal{D}\hat{r}^{\frac{2^t-1}{2^t}}$ by induction using $\mathcal{L}_t = \sqrt{\mathcal{L}_{t-1} \mathcal{D} \hat r}$. Then
	$$\frac{p_t^D}{p_{t-1}^D} = \sqrt{\frac{\mathcal{L}_{t-1}}{\mathcal{D}\hat r}}
	= \sqrt{\frac{\mathcal{D}\hat{r}^{\frac{2^{t-1}-1}{2^{t-1}}}}{\mathcal{D}\hat r}}
	= \hat{r}^{\frac{1}{2}\Big(\frac{2^{t-1}-1}{2^{t-1}}-1\Big)} = \hat{r}^{-2^{-t}}.$$
	And so $\ln \frac{p_t^D}{p_{t-1}^D} = -2^{-t} \ln \hat r$.
	
	Next, as $\bar\mu_t = (1-\delta)\bar\mu_{t-1} + \delta \ln \frac{p_t^D}{p_{t-1}^D}$, it is straightforward to verify by induction that
	$$\bar\mu_t = (1-\delta)^t \bar\mu_0 - \delta \ln \hat r \sum_{k=1}^t 2^{-k}(1-\delta)^{t-k}.$$
	
	\paragraph{Case I:} $\delta = 1/2$. The series in $\bar\mu_t$ becomes
	$$\sum_{k=1}^t 2^{-k}(1-\delta)^{t-k} = \sum_{k=1}^t 2^{-k} 2^{-(t-k)}
	= \sum_{k=1}^t 2^{-t} = \frac{t}{2^t}.$$
	Then we have
	$\bar\mu_t = 2^{-t}\Big( \bar\mu_0 - \frac{1}{2}t \ln \hat r\Big)$.
	The first term $\rightarrow 0$ since $0\leq \delta < 1$. The second term $\rightarrow 0$ by L'Hopital's rule. Thus $\bar\mu_t \rightarrow 0$ as $t\rightarrow \infty$.
	
	The contributing term to volatility at time $t$, after substituting and simplifying terms, is
	$$\ln \frac{p_t^D}{p_{t-1}^D} - \bar\mu_t
	= \frac{t/2-1}{2^t}\ln \hat r - 2^{-t} \bar\mu_0.$$
	Then DStablecoin volatility evolves according to
	$$\begin{aligned}
	\bar\sigma_t^2 &= (1-\delta)\bar\sigma_{t-1}^2 + \delta\Big(\ln \frac{p^D_t}{p^D_{t-1}} - \bar\mu_t\Big)^2 \\
	&= \sum_{k=1}^t (1-\delta)^{t-k} \delta \Big(\ln \frac{p_k^D}{p_{k-1}^D} -\bar\mu_k\Big)^2 + (1-\delta)^t \bar\sigma_0^2 \\
	&= \sum_{k=1}^t 2^{-(t-k)} \delta \Big( \frac{k/2-1}{2^k}\ln \hat r - 2^{-k} \bar\mu_0 \Big)^2 + 2^{-t} \bar\sigma_0^2 \\
	&= \sum_{k=1}^t 2^{-(t-k)} \delta 2^{-2k} \Big( (k/2-1)\ln \hat r - \bar\mu_0\Big)^2 + 2^{-t} \bar\sigma_0^2 \\
	&= 2^{-t} \sum_{k=1}^t 2^{-k-1} \Big( (k/2-1)\ln \hat r - \bar\mu_0\Big)^2 + 2^{-t} \bar\sigma_0^2. \\
	\end{aligned}$$
	The second line follows from straightforward induction. As $t\rightarrow\infty$, the series converges from exponential decay. Then both terms $\rightarrow 0$ because of the factor of $2^{-t}$. Thus $\bar\sigma_t^2 \rightarrow 0$.

	\paragraph{Case II:} $\delta \neq 1/2$. The series in $\bar\mu_t$ is a geometric progression
	$$\begin{aligned}
	\sum_{k=1}^t 2^{-k}(1-\delta)^{t-k} &= \sum_{k=1}^t (1-\delta)^t \Big(2(1-\delta)\Big)^{-k} \\
	&= \frac{(1-\delta)^t\Big( 2(1-\delta)^{-1} - 2^{-t-1}(1-\delta)^{-t-1}\Big)}{1- 2(1-\delta)^{-1}} \\
	&= \frac{(1-\delta)^t - 2^{-t}}{2(1-\delta)-1}
	\end{aligned}$$
	Then we have
	$\bar\mu_t = (1-\delta)^t \bar\mu_0 - \delta \frac{(1-\delta)^t-2^{-t}}{2(1-\delta)-1} \ln \hat r$, which converges to 0 as $t\rightarrow\infty$.
	
	The contributing term to volatility at time $t$, after substituting and simplifying terms, is
	$$\ln \frac{p_t^D}{p_{t-1}^D} - \bar\mu_t
	= (1-\delta)^t \bar\mu_0 - \frac{(1-\delta)^t -2^{-t+1}(1-\delta)}{2(1-\delta)-1}\ln \hat r.$$
	The DStablecoin volatility evolves according to
	$$\begin{aligned}
	\bar\sigma_t^2 &= \sum_{k=1}^t (1-\delta)^{t-k} \delta \Big(\ln \frac{p_k^D}{p_{k-1}^D} -\bar\mu_k\Big)^2 + (1-\delta)^t \bar\sigma_0^2 \\
	&= \sum_{k=1}^t (1-\delta)^{t-k}\delta \Big( (1-\delta)^k \bar\mu_0 - \frac{(1-\delta)^k -2^{-k+1}(1-\delta)}{2(1-\delta)-1}\ln \hat r \Big)^2 + (1-\delta)^t \bar\sigma_0^2. \\
	\end{aligned}$$
	Note that because $(1-\delta) \geq 1/2$, we have
	$$\begin{aligned}
	|(1-\delta)^t - 2^{-t+1}(1-\delta)| &\leq (1-\delta)^t + 2^{-t+1}(1-\delta) \\
	&\leq 2(1-\delta)^t.
	\end{aligned}$$
	Thus we have
	$$\begin{aligned}
	\bar\sigma_t^2 &\leq (1-\delta)^t \sum_{k=1}^t \frac{\delta}{(1-\delta){^k}} \Big( (1-\delta)^k \bar\mu_0 + \frac{2(1-\delta)^k}{2(1-\delta)-1}\ln \hat r \Big)^2 + (1-\delta)^t \bar\sigma_0^2 \\
	&= (1-\delta)^t \sum_{k=1}^t (1-\delta)^{k}\delta \Big( \bar\mu_0 + \frac{2}{2(1-\delta)-1}\ln \hat r \Big)^2 + (1-\delta)^t \bar\sigma_0^t.
	\end{aligned}$$
	As $t\rightarrow\infty$, the series converges from exponential decay. Then both terms $\rightarrow 0$ because of the factor of $(1-\delta)^t$. Thus $\bar\sigma_t^2 \rightarrow 0$.
\end{proof}


\end{document}